\documentclass[aps,pre,reprint,longbibliography,floatfix]{revtex4-2}

\usepackage[utf8]{inputenc}
\usepackage[T1]{fontenc}
\usepackage{amsmath,amssymb,latexsym,graphicx}
\usepackage{newtxtext,newtxmath}
\usepackage{bbold,anyfontsize}
\usepackage[dvipsnames]{xcolor}
\usepackage[
  colorlinks=true,
  urlcolor=BlueViolet,
  citecolor=BlueViolet,
  filecolor=BlueViolet,
  linkcolor=BlueViolet
]{hyperref}

\begin{document}

\title{
  Conditioning the complexity of random landscapes on marginal optima
}

\author{Jaron Kent-Dobias}
\email{jaron.kent-dobias@roma1.infn.it}
\affiliation{
  Istituto Nazionale di Fisica Nucleare, Sezione di Roma I, Rome, Italy 00184
}

\begin{abstract}
  Marginal optima are minima or maxima of a function with many nearly flat
  directions. In settings with many competing optima, marginal ones tend to
  attract algorithms and physical dynamics. Often, the important family of
  marginal attractors are a vanishing minority compared with nonmarginal optima
  and other unstable stationary points. We introduce a generic technique for
  conditioning the statistics of stationary points in random landscapes on
  their marginality, and apply it in three isotropic settings with
  qualitatively different structure: in the spherical spin-glasses, where the
  energy is Gaussian and its Hessian is GOE; in multispherical spin glasses,
  which are Gaussian but non-GOE; and in sums of squared spherical random
  functions, which are non-Gaussian. In these problems we are able to fully
  characterize the distribution of marginal optima in the landscape, including
  when they are in the minority.
\end{abstract}

\maketitle

\section{Introduction}

Systems with rugged landscapes are important across many disciplines, from the
physics of glasses and spin-glasses to statistical inference problems \cite{Ros_2023_High-dimensional}. The
behavior of these systems is best understood when equilibrium or optimal
solutions are studied and weighted averages can be taken statically over all
possible configurations. However, such systems are also infamous for their
tendency to defy equilibrium and optimal expectations in practice, due to the
presence of dynamic transitions or crossovers that leave physical or
algorithmic dynamics stuck exploring only a subset of configurations \cite{Biroli_2013_Perspective, Krzakala_2007_Landscape}.

In mean-field settings, it was long thought that physical and many algorithmic
dynamics would get stuck at a specific energy level, called the threshold
energy. The threshold energy is the energy level at which level sets of the
landscape transition from containing mostly saddle points to containing mostly
minima. The level set associated with this threshold energy contains mostly \emph{marginal
minima}, or minima whose Hessian matrix have a continuous spectral density over
all sufficiently small positive eigenvalues. In most circumstances the spectrum
is \emph{pseudogapped}, which means that the spectral density smoothly
approaches zero as zero eigenvalue is approached from above.

However, recent work found that the threshold energy is not important even for
simple gradient descent dynamics \cite{Folena_2020_Rethinking, Folena_2023_On, ElAlaoui_2020_Algorithmic}.
Depending on the initial condition of the system and the nature of the
dynamics, the energy reached can be above or below the threshold energy, while
in some models the threshold energy is completely inaccessible to any dynamics
\cite{Kent-Dobias_2023_How}. Though it is still not known how to predict the
energy level that many simple algorithms will reach, the results all share one
commonality: the minima found are still marginal, despite being in the minority
compared to stiff minima or saddle points. This ubiquity of behavior suggests
that the distribution of marginal minima can be used to bound out-of-equilibrium dynamical
behavior.

Despite their importance in a wide variety of in and out of equilibrium
settings \cite{Muller_2015_Marginal, Anderson_1984_Lectures,
Sommers_1984_Distribution, Parisi_1995-01_On, Horner_2007_Time,
Pankov_2006_Low-temperature, Erba_2024_Quenches, Efros_1985_Coulomb,
Shklovskii_2024_Half, Folena_2022_Marginal}, it is not straightforward to condition on the
marginality of minima using the traditional methods for analyzing the
distribution of minima in rugged landscapes. Using the method of a Legendre
transformation of the Parisi parameter corresponding to a set of real replicas,
one can force the result to correspond with marginal minima by tuning the value
of that parameter \cite{Monasson_1995_Structural}. However, this results in
only a characterization of the threshold energy and cannot characterize
marginal minima at other energies where they are a minority.

The alternative approach, used to great success in the spherical spin glasses, is to
start by making a detailed understanding of the Hessian matrix at stationary
points. Then, one can condition the analysis on whatever properties of the
Hessian are necessary to lead to marginal minima. This strategy is so
successful in the spherical spin glasses because it is straightforward to implement.
First, the shape of the Hessian's spectrum is independent of energy and even
whether one sits at a stationary point or not. This is a property of models
whose energy is a Gaussian random variable \cite{Fyodorov_2004_Complexity, Bray_2007_Statistics}.
Furthermore, a natural parameter in the analysis of these models linearly
shifts the spectrum of the Hessian. Therefore, tuning this parameter to a
specific constant value allows one to require that the Hessian spectrum have a
pseudogap, and therefore that the associated minima be marginal. Unfortunately
this strategy is less straightforward to generalize to other models. Many models of interest,
especially in inference problems, have Hessian statistics that are poorly
understood. This is especially true for the statistics of the Hessian
conditioned to lie at stationary points, which is necessary to understand in
models whose energy is non-Gaussian.

Here, we introduce a generic method for conditioning the statistics of
stationary points on their marginality. The technique makes use of a novel way
to condition an integration measure to select only configurations that result
in a certain value of the smallest eigenvalue of a matrix. By requiring that
the smallest eigenvalue of the Hessian at stationary points be zero, and
further looking for a sign that the zero eigenvalue lies at the edge of a
continuous spectrum, we enforce the condition that the spectrum has a
pseudogap, and is therefore marginal. We demonstrate the method on the
spherical spin glasses, where it is unnecessary but instructive, and on extensions of
the spherical models where the technique is more useful.
In a related work, we compare the marginal complexity with the performance
of gradient descent and approximate message passing algorithms \cite{Kent-Dobias_2024_Algorithm-independent}.

An outline of this paper follows. In Section \ref{sec:eigenvalue} we introduce the technique for conditioning on
the smallest eigenvalue and how to extend it to further condition on the
presence of a pseudogap. We provide a simple but illustrative example using a
GOE matrix with a shifted diagonal. In Section \ref{sec:marginal.complexity} we apply this
technique to the problem of characterizing marginal minima in random
landscapes. The following Section \ref{sec:examples} gives several examples of
the marginal complexity applied to specific models of increasing difficulty.
Finally, Section \ref{sec:conclusion} summarizes this work and suggests
necessary extensions.

\section{Conditioning on the smallest eigenvalue}
\label{sec:eigenvalue}

In this section, we introduce a general method for conditioning a measure on
the smallest eigenvalue of some matrix that depends on it. In Section
\ref{sec:shifted.GOE} we show how this works in perhaps the simplest example of
GOE random matrices with a shifted diagonal. In the final subsection we
describe how to extend this method to condition on the presence of a pseudogap
at the bottom on the spectrum.

\subsection{The general method}

Consider an $N\times N$ real symmetric matrix $A$. An arbitrary function $g$ of the
minimum eigenvalue of $A$ can be expressed using integrals over $\mathbf
s\in\mathbb R^N$ as
\begin{equation} \label{eq:λmin}
  g(\lambda_\textrm{min}(A))
  =\lim_{\beta\to\infty}\int
    \frac{d\mathbf s\,\delta(N-\|\mathbf s\|^2)e^{-\beta\mathbf s^TA\mathbf s}}
      {\int d\mathbf s'\,\delta(N-\|\mathbf s'\|^2)e^{-\beta\mathbf s'^TA\mathbf s'}}
    g\left(\frac{\mathbf s^TA\mathbf s}N\right)
\end{equation}
In the limit of large $\beta$, each integral concentrates among vectors
$\mathbf s$ in the eigenspace of $A$ corresponding to the smallest eigenvalue
of $A$. This produces
\begin{equation}
  \begin{aligned}
    &\lim_{\beta\to\infty}\int\frac{
      d\mathbf s\,\delta(N-\|\mathbf s\|^2)e^{-\beta\mathbf s^TA\mathbf s}
    }{
      \int d\mathbf s'\,\delta(N-\|\mathbf s'\|^2)e^{-\beta\mathbf s'^TA\mathbf s'}
    }g\left(\frac{\mathbf s^TA\mathbf s}N\right) \\
    &=\int\frac{
      d\mathbf s\,\delta(N-\|\mathbf s\|^2)\mathbb 1_{\operatorname{ker}(A-\lambda_\mathrm{min}(A)I)}(\mathbf s)
    }{
      \int d\mathbf s'\,\delta(N-\|\mathbf s'\|^2)\mathbb 1_{\operatorname{ker}(A-\lambda_\mathrm{min}(A)I)}(\mathbf s')}g\left(\frac{\mathbf s^TA\mathbf s}N\right) \\
    &=g(\lambda_\mathrm{min}(A))
    \frac{\int d\mathbf s\,\delta(N-\|\mathbf s\|^2)\mathbb 1_{\operatorname{ker}(A-\lambda_\mathrm{min}(A)I)}(\mathbf s)}{\int d\mathbf s'\,\delta(N-\|\mathbf s'\|^2)\mathbb 1_{\operatorname{ker}(A-\lambda_\mathrm{min}(A)I)}(\mathbf s')} \\
    &=g(\lambda_\mathrm{min}(A))
  \end{aligned}
\end{equation}
as desired.
The first relation extends a technique for calculating the typical minimum eigenvalue of an ensemble of matrices first introduced by
\citeauthor{Ikeda_2023_Bose-Einstein-like} and later used by
\citeauthor{Kent-Dobias_2024_Arrangement} in the context of random landscapes, and is similar to an earlier technique for conditioning the value of the ground state energy in random landscapes by \citeauthor{Fyodorov_2013_Topology} \cite{Ikeda_2023_Bose-Einstein-like, Kent-Dobias_2024_Algorithm-independent, Fyodorov_2013_Topology, Fyodorov_2018_Hessian}. A Boltzmann distribution is introduced
over a spherical model whose Hamiltonian is quadratic with interaction matrix
given by $A$. In the limit of zero temperature, the measure will concentrate on
the ground states of the model, which correspond with the eigenspace of $A$
associated with its minimum eigenvalue $\lambda_\mathrm{min}$. The second
relation uses the fact that, once restricted to the sphere $\|\mathbf
s\|^2=N$ and the minimum eigenspace, $\mathbf s^TA\mathbf s=\mathbf s^T\mathbf s\lambda_\mathrm{min}(A)=N\lambda_\mathrm{min}(A)$.

The relationship is formal, but we can make use of the fact that the integral
expression with a Gibbs distribution can be manipulated with replica
techniques, averaged over, and in general treated with a physicist's toolkit.
In particular, we have specific interest in using
$g(\lambda_\mathrm{min}(A))=\delta(\lambda_\mathrm{min}(A))$, a Dirac
delta-function, which can be inserted into averages over ensembles of matrices
$A$ (or indeed more complicated averages) in order to condition that the
minimum eigenvalue is zero.

\subsection{Simple example: shifted GOE}
\label{sec:shifted.GOE}

We demonstrate the efficacy of the technique by rederiving a well-known result:
the large-deviation function for pulling an eigenvalue from the bulk of the
GOE spectrum.
Consider an ensemble of $N\times N$ matrices $A=B+\mu I$ for $B$ drawn from the GOE ensemble with entries
whose variance is $\sigma^2/N$. We know that the bulk spectrum of $A$ is a
Wigner semicircle with radius $2\sigma$ shifted by a constant $\mu$.
Therefore, for $\mu=2\sigma$, the minimum eigenvalue will typically be zero,
while for $\mu>2\sigma$ the minimum eigenvalue would need to be a large
deviation from the typical spectrum and its likelihood will be exponentially
suppressed with $N$. For $\mu<2\sigma$, the bulk of the typical spectrum contains
zero and therefore a larger $N^2$ deviation, moving an extensive number of
eigenvalues, would be necessary \cite{Dean_2006_Large}. This final case cannot be quantified by this
method, but instead the nonexistence of a large deviation linear in $N$ appears
as the emergence of an imaginary part in the large deviation function.

\begin{figure}
  \hspace{1.3em}
  \includegraphics{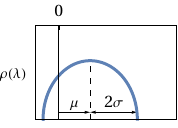}
  \hspace{-2em}
  \includegraphics{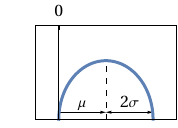}
  \hspace{-2em}
  \includegraphics{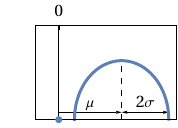}
  \\
  \includegraphics{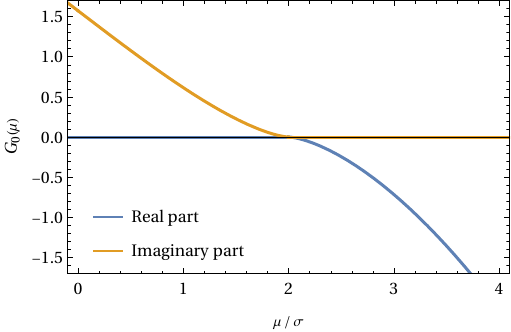}
  \caption{
    The large deviation function $G_0(\mu)$ defined in
    \eqref{eq:large.dev} as a function of the shift $\mu$ to the
    GOE diagonal. $G_0(2\sigma)=0$, while for
    $\mu>2\sigma$ it is negative and for $\mu<2\sigma$ it gains an
    imaginary part. The top panels show schematically what happens to the
    spectral density in each of these regimes. For $\mu<2\sigma$, an $N^2$
    large deviation would be required to fix the smallest eigenvalue to zero
    and the calculation breaks down, leading to the imaginary part. For
    $\mu>2\sigma$ the spectrum can satisfy the constraint on the smallest
    eigenvalue by isolating a single eigenvalue at zero at the cost of an
    order-$N$ large deviation. At the transition point $\mu=2\sigma$ the
    spectrum is pseudogapped.
  } \label{fig:large.dev}
\end{figure}

To compute this large deviation function, we will employ the method outlined in
the previous subsection to calculate
\begin{equation} \label{eq:large.dev}
  \begin{aligned}
    e^{NG_{\lambda^*}(\mu)}
    &=P\big(\lambda_\mathrm{min}(B+\mu I)=\lambda^*\big) \\
    &=\overline{\delta\big(N\lambda^*-N\lambda_\mathrm{min}(B+\mu I)\big)}
  \end{aligned}
\end{equation}
where the overline is the average over $B$, and we have defined the large
deviation function $G_{\lambda^*}(\mu)$.
Using the representation of $\lambda_\mathrm{min}$ defined in \eqref{eq:λmin}, we have
\begin{widetext}
\begin{equation}
  e^{NG_{\lambda^*}(\mu)}
  =\overline{
    \lim_{\beta\to\infty}\int\frac{d\mathbf s\,\delta(N-\|\mathbf s\|^2)e^{-\beta\mathbf s^T(B+\mu I)\mathbf s}}
    {\int d\mathbf s'\,\delta(N-\|\mathbf s'\|^2)e^{-\beta\mathbf s'^T(B+\mu I)\mathbf s'}}\,\delta\big(N\lambda^*-\mathbf s^T(B+\mu I)\mathbf s\big)
  }
\end{equation}
Using replicas to treat the denominator ($x^{-1}=\lim_{m\to0}x^{m-1}$)
and transforming the $\delta$-function to its Fourier
representation, we have
\begin{equation}
  e^{NG_{\lambda^*}(\mu)}
  =\overline{\lim_{\beta\to\infty}\lim_{m\to0}\int d\hat\lambda\prod_{\alpha=1}^m\left[d\mathbf s^\alpha\,\delta(N-\|\mathbf s^\alpha\|^2)\right]
  \exp\left\{-\beta\sum_{\alpha=1}^m(\mathbf s^\alpha)^T(B+\mu I)\mathbf s^\alpha+\hat\lambda\left[N\lambda^*-(\mathbf s^1)^T(B+\mu I)\mathbf s^1\right]\right\}}
\end{equation}
having introduced the auxiliary parameter $\hat\lambda$ in the Fourier representation of
the $\delta$-function. The whole expression, so transformed, is an
exponential integral linear in the matrix $B$. Taking the average over $B$, we
find
\begin{equation}
  \begin{aligned}
  &e^{NG_{\lambda^*}(\mu)}
  =\lim_{\beta\to\infty}\lim_{m\to0}\int d\hat\lambda\prod_{\alpha=1}^m\left[d\mathbf s^\alpha\,\delta(N-\|\mathbf s^\alpha\|^2)\right] \\
  &\hspace{10em}
  \times\exp\left\{N\left[\hat\lambda(\lambda^*-\mu)-m\beta\mu\right]+\frac{\sigma^2}{N}\left[\beta^2\sum_{\alpha\gamma}^m(\mathbf s^\alpha\cdot\mathbf s^\gamma)^2
    +2\beta\hat\lambda\sum_\alpha^m(\mathbf s^\alpha\cdot\mathbf s^1)^2
    +\hat\lambda^2N^2
  \right]\right\}
  \end{aligned}
\end{equation}
\end{widetext}
We make the Hubbard--Stratonovich transformation to the matrix field
$Q^{\alpha\beta}=\frac1N\mathbf s^\alpha\cdot\mathbf s^\beta$. This produces an integral expression of the form
\begin{equation}
  e^{NG_{\lambda^*}(\mu)}
  =\lim_{\beta\to\infty}\lim_{m\to0}\int d\hat\lambda\,dQ\,
  e^{N\mathcal U_\mathrm{GOE}(\hat\lambda,Q\mid\beta,\lambda^*,\mu)}
\end{equation}
where the effective action $\mathcal U_\mathrm{GOE}$ is given by
\begin{equation} \label{eq:goe.action}
  \begin{aligned}
    &\mathcal U_\textrm{GOE}(\hat\lambda, Q\mid\beta,\lambda^*,\mu)
    =\hat\lambda(\lambda^*-\mu)+\lim_{m\to0}\bigg\{-m\beta\mu \\
    &+\sigma^2\left[\beta^2\sum_{\alpha\gamma}^m(Q^{\alpha\gamma})^2
        +2\beta\hat\lambda\sum_\alpha^m(Q^{1\alpha})^2
      +\hat\lambda^2
    \right]+\frac12\log\det Q\bigg\}
  \end{aligned}
\end{equation}
and $Q^{\alpha\alpha}=1$ because of the spherical constraint. We can evaluate this
integral using the saddle point method. We make a replica symmetric ansatz for
$Q$, because this is a 2-spin spherical model, but with the first row singled out because
of its unique coupling with $\hat\lambda$. The resulting matrix has the form
\begin{equation} \label{eq:Q.structure}
  Q=\begin{bmatrix}
    1&\tilde q_0&\tilde q_0&\cdots&\tilde q_0\\
    \tilde q_0&1&q_0&\cdots&q_0\\
    \tilde q_0&q_0&1&\ddots&q_0\\
    \vdots&\vdots&\ddots&\ddots&\vdots\\
    \tilde q_0&q_0&q_0&\cdots&1
  \end{bmatrix}
\end{equation}
The relevant expressions in the effective action produce
\begin{align}
  &\sum_{\alpha\beta}(Q^{\alpha\beta})^2=m+2(m-1)\tilde q_0^2+(m-1)(m-2)q_0^2 \\
  &\sum_\alpha(Q^{1\alpha})^2=1+(m-1)\tilde q_0^2 \\
  &\begin{aligned}
    &\log\det Q=(m-2)\log(1-q_0) \\
    &\hspace{6em}+\log\big[1+(m-2)q_0-(m-1)\tilde q_0^2\big]
  \end{aligned}
\end{align}
Inserting these expressions into the effective action and taking the limit of
$m$ to zero, we arrive at
\begin{equation}
  e^{NG_{\lambda^*}(\mu)}
  =\lim_{\beta\to\infty}\int d\hat\lambda\,dq_0\,d\tilde q_0\,
  e^{N\mathcal U_\textrm{GOE}(\hat\lambda,q_0,\tilde q_0\mid\beta,\lambda^*,\mu)}
\end{equation}
with the new effective action
\begin{equation}
  \begin{aligned}
    &\mathcal U_\mathrm{GOE}(\hat\lambda,q_0,\tilde q_0\mid\beta,\lambda^*,\mu) \\
    &\quad=\hat\lambda(\lambda^*-\mu)+\sigma^2\left[
      2\beta^2(q_0^2-\tilde q_0^2)+2\beta\hat\lambda(1-\tilde q_0^2)+\hat\lambda^2
    \right] \\
    &\qquad-\log(1-q_0)+\frac12\log(1-2q_0+\tilde q_0^2)
  \end{aligned}
\end{equation}
We need to evaluate the integral above using the saddle point method, but in the limit of $\beta\to\infty$.
We expect the overlaps to concentrate on one as $\beta$ goes to infinity. We therefore take
\begin{align}
  \label{eq:q0.limit}
  q_0&=1-y\beta^{-1}-z\beta^{-2}+O(\beta^{-3})
  \\
  \label{eq:q0t.limit}
  \tilde q_0&=1-\tilde y\beta^{-1}-(z+\Delta z)\beta^{-2}+O(\beta^{-3})
\end{align}
However, taking the limit with $y\neq\tilde y$ results in an expression for the
action that diverges with $\beta$. To cure this, we must take $\tilde y=y$. The result is
\begin{equation}
  \begin{aligned}
    &\mathcal U_\textrm{GOE}(\hat\lambda,y,\Delta z\mid\infty,\lambda^*,\mu)
    =\hat\lambda(\lambda^*-\mu) \\
    &\qquad+\sigma^2\big[
      \hat\lambda^2+4(y+\Delta z)
    \big]
    +\frac12\log\left(1-\frac{2\Delta z}{y^2}\right)
  \end{aligned}
\end{equation}
Extremizing this action over the new parameters $y$, $\Delta z$, and $\hat\lambda$, we find
\begin{align}
  \hat\lambda&=\frac1\sigma\sqrt{\left(\frac{\mu-\lambda^*}{2\sigma}\right)^2-1}
  \\
  y&=\frac1{2\sigma}\left[
    \frac{\mu-\lambda^*}{2\sigma}+\sqrt{\left(\frac{\mu-\lambda^*}{2\sigma}\right)^2-1}
  \right]^{-1}
  \\
  \Delta z&=\frac1{4\sigma^2}\left[
    \left(\frac{\mu-\lambda^*}{2\sigma}\right)^2-1
    -\frac{\mu-\lambda^*}{2\sigma}\sqrt{\left(\frac{\mu-\lambda^*}{2\sigma}\right)^2-1}
  \right]
\end{align}
Inserting this solution into the effective action we arrive at
\begin{equation} \label{eq:goe.large.dev}
  \begin{aligned}
    &G_{\lambda^*}(\mu)
    =\mathop{\textrm{extremum}}_{\hat\lambda,y,\Delta z}
    \mathcal U_\mathrm{GOE}(\hat\lambda,y,\Delta z\mid\infty,\lambda^*,\mu) \\
    &=-\frac{\mu-\lambda^*}{2\sigma}\sqrt{\left(\frac{\mu-\lambda^*}{2\sigma}\right)^2-1} \\
    &\hspace{5em}-\log\left[
      \frac{\mu-\lambda^*}{2\sigma}-\sqrt{\left(\frac{\mu-\lambda^*}{2\sigma}\right)^2-1}
    \right]
  \end{aligned}
\end{equation}
This function is plotted in Fig.~\ref{fig:large.dev} for $\lambda^*=0$. For $\mu<2\sigma$, $G_{0}(\mu)$ has an
imaginary part. This indicates that the existence of a zero minimum eigenvalue when $\mu<2\sigma$ corresponds with a large deviation that grows faster than $N$,
rather like $N^2$, since in this regime the bulk of the typical spectrum is
over zero and therefore extensively many eigenvalues have to have large
deviations in order for the smallest eigenvalue to be zero \cite{Dean_2006_Large}. For
$\mu\geq2\sigma$ this function gives the large deviation function for the
probability of seeing a zero eigenvalue given the shift $\mu$.
$\mu=2\sigma$ is the maximum of the function with a real value, and
corresponds to the intersection of the typical bulk spectrum with zero, i.e., a
pseudogap.

Here, we see what appears to be a general heuristic for identifying the saddle
parameters for which the spectrum is pseudogapped: the equivalent of this
large-deviation function will lie on the singular boundary between a purely
real and complex value.

\subsection{Conditioning on a pseudogap}
\label{sec:pseudogap}

We have seen that this method effectively conditions a random matrix ensemble
on its lowest eigenvalue being zero. However, this does not correspond on its
own to marginality. In the previous example, most values of $\mu$ where
the calculation was valid correspond to matrices with a single isolated
eigenvalue. However, the marginal minima we are concerned with have
pseudogapped spectra, where the continuous part of the spectral density has a
lower bound at zero.

Fortunately, our calculation can be modified to ensure that we consider only
pseudogapped spectra. First, we insert a shift $\mu$ by hand into the `natural'
spectrum of the problem at hand, conditioning the trace to have a specific
value $\mu=\frac1N\operatorname{Tr}A$. Then, we choose this artificial shift so that
the resulting conditioned spectra are pseudogapped. As seen the previous
subsection, this can be done by starting from a sufficiently large $\mu$ and
decreasing it until the calculation develops an imaginary part, signaling the
breakdown of the large-deviation principle at order $N$.

In isotropic or zero-signal landscapes, there is another way to condition on a
pseudogap. In such landscapes, the typical spectrum does not have an isolated
eigenvalue. Therefore, for a given $\mu$ the bottom of the spectrum can be located by looking for the value $\lambda^*$ that maximizes the (real) large deviation function.
Inverting this reasoning, we can find the value $\mu=\mu_\textrm m$
corresponding to a marginal spectrum by requiring that the large deviation
function has a maximum in $\lambda^*$ at $\lambda^*=0$, or
\begin{equation}
  0=\frac\partial{\partial\lambda^*}G_{\lambda^*}(\mu_\mathrm m)\bigg|_{\lambda^*=0}
\end{equation}
In the example problem of section \ref{sec:shifted.GOE}, this corresponds precisely to $\mu_\mathrm m=2\sigma$,
the correct marginal shift. Note that when we treat the Dirac $\delta$ function
using its Fourier representation with auxiliary parameter $\hat\lambda$, as in
the previous subsection, this condition corresponds with choosing $\mu$ such
that $\hat\lambda=0$.

\section{Marginal complexity in random landscapes}
\label{sec:marginal.complexity}

The methods of the previous section can be used in diverse settings. However,
we are interested in applying them to study stationary points in random
landscapes whose Hessian spectrum has a pseudogap -- that is, that are
marginal. In Section \ref{sec:marginal.kac-rice} we define the marginal
complexity using the tools of the previous section. In Section
\ref{sec:general.features} we review several general features in a physicists'
approach to computing the marginal complexity. In Section
\ref{sec:superspace_kac-rice} we introduce a representation of the marginal
complexity in terms of an integral over a superspace, which condenses the
notation and the resulting calculation and which we will use in one of our
examples in the next section.

\subsection{Marginal complexity from Kac--Rice}
\label{sec:marginal.kac-rice}

The situation in the study of random landscapes is often as follows: an
ensemble of smooth energy functions $H:\mathbb R^N\to\mathbb R$ defines a family of random
landscapes, often with their configuration space subject to one or more
constraints of the form $g(\mathbf x)=0$ for $\mathbf x\in\mathbb R^N$. The
typical geometry of landscapes drawn from the ensemble is studied by their complexity, or the average
logarithm of the number of stationary points with certain properties, e.g., of
marginal minima at a given energy.

Such problems can be studied using the method of Lagrange multipliers, with one
introduced for every constraint. If the configuration space is defined by $r$
constraints, then the problem of identifying stationary points is reduced to
extremizing the Lagrangian
\begin{equation}
  L(\mathbf x,\pmb\omega)=H(\mathbf x)+\sum_{i=1}^r\omega_ig_i(\mathbf x)
\end{equation}
with respect to $\mathbf x$ and the Lagrange multipliers
$\pmb\omega=\{\omega_1,\ldots,\omega_r\}$. To write the gradient and Hessian of the energy, which are necessary to count stationary points, care must be taken to ensure they are constrained to the tangent space of the configuration manifold. For our purposes, the Lagrangian formalism offers a solution: the gradient $\nabla H:\mathbb R^N\times\mathbb R^r\to\mathbb R^N$ and
Hessian $\operatorname{Hess} H:\mathbb R^N\times\mathbb R^r\to\mathbb R^{N\times N}$ of the energy $H$ can be written as the simple vector derivatives of
the Lagrangian $L$, with
\begin{align}
  &\nabla H(\mathbf x,\pmb\omega)
  =\partial L(\mathbf x,\pmb\omega)
  =\partial H(\mathbf x)+\sum_{i=1}^r\omega_i\partial g_i(\mathbf x)
  \\
  &\begin{aligned}
    \operatorname{Hess}H(\mathbf x,\pmb\omega)
    &=\partial\partial L(\mathbf x,\pmb\omega) \\
    &=\partial\partial H(\mathbf x)+\sum_{i=1}^r\omega_i\partial\partial g_i(\mathbf x)
  \end{aligned}
\end{align}
where $\partial=\frac\partial{\partial\mathbf x}$ will always represent the
derivative with respect to the vector argument $\mathbf x$. Note that unlike
the energy, which is a function of the configuration $\mathbf x$ alone, the
gradient and Hessian depend also on the Lagrange multipliers $\pmb\omega$. In situations
with an extensive number of constraints, it is important to take seriously
contributions of the form $\frac{\partial^2L}{\partial\mathbf
x\partial\pmb\omega}$ to the Hessian \cite{Kent-Dobias_2024_On}. However, the cases we study here have
$N^0$ constraints and these contributions appear as finite-$N$ corrections.

The number of
stationary points in a landscape for a particular function $H$ is found by
integrating over the Kac--Rice measure
\begin{equation} \label{eq:kac-rice.measure}
  \begin{aligned}
    &d\nu_H(\mathbf x,\pmb\omega) \\
    &\quad=
    d\mathbf x\,d\pmb\omega\,\delta\big(\mathbf g(\mathbf x)\big)
    \,\delta\big(\nabla H(\mathbf x,\pmb\omega)\big)
    \,\big|\det\operatorname{Hess}H(\mathbf x,\pmb\omega)\big|
  \end{aligned}
\end{equation}
with a $\delta$-function of the gradient and the constraints ensuring that we
count valid stationary points, and the determinant of the Hessian serving as
the Jacobian of the argument to the $\delta$ function \cite{Kac_1943_On, Rice_1939_The}. It is usually more
interesting to condition the count on interesting properties of the stationary
points, like the energy and spectrum trace, or
\begin{equation} \label{eq:kac-rice.measure.2}
  \begin{aligned}
    &d\nu_H(\mathbf x,\pmb\omega\mid E,\mu) \\
    &=d\nu_H(\mathbf x,\pmb\omega)\,
    \delta\big(NE-H(\mathbf x)\big)
    \,\delta\big(N\mu-\operatorname{Tr}\operatorname{Hess}H(\mathbf x,\pmb\omega)\big)
  \end{aligned}
\end{equation}
We specifically want to control the value of the minimum eigenvalue of the Hessian
at the stationary points. Using the method introduced in Section
\ref{sec:eigenvalue}, we can write the number of stationary points with energy
$E$, Hessian trace $\mu$, and smallest eigenvalue $\lambda^*$ as
\begin{widetext}
\begin{equation}
  \begin{aligned}
    &\mathcal N_H(E,\mu,\lambda^*)
    =\int d\nu_H(\mathbf x,\pmb\omega\mid E,\mu)\,\delta\big(N\lambda^*-\lambda_\mathrm{min}(\operatorname{Hess}H(\mathbf x,\pmb\omega))\big) \\
    &=\lim_{\beta\to\infty}\int d\nu_H(\mathbf x,\pmb\omega\mid E,\mu)
    \frac{d\mathbf s\,\delta(N-\|\mathbf s\|^2)\delta(\mathbf s^T\partial\mathbf g(\mathbf x))e^{-\beta\mathbf s^T\operatorname{Hess}H(\mathbf x,\pmb\omega)\mathbf s}}
    {\int d\mathbf s'\,\delta(N-\|\mathbf s'\|^2)\delta(\mathbf s'^T\partial\mathbf g(\mathbf x))e^{-\beta\mathbf s'^T\operatorname{Hess}H(\mathbf x,\pmb\omega)\mathbf s'}}
    \delta\big(N\lambda^*-\mathbf s^T\operatorname{Hess}H(\mathbf x,\pmb\omega)\mathbf s\big)
  \end{aligned}
\end{equation}
where the additional $\delta$-functions
\begin{equation}
  \delta(\mathbf s^T\partial\mathbf g(\mathbf x))
  =\prod_{s=1}^r\delta(\mathbf s^T\partial g_i(\mathbf x))
\end{equation}
ensure that the integrals involving potential eigenvectors $\mathbf s$ are constrained to
the tangent space of the configuration manifold at the point $\mathbf x$.

The
complexity of points with a specific energy, stability, and minimum eigenvalue
is defined as the average over the ensemble of functions $H$ of the logarithm
of the number $\mathcal N_H$ of stationary points, or
\begin{equation}
  \Sigma_{\lambda^*}(E,\mu)
  =\frac1N\overline{\log\mathcal N_H(E,\mu,\lambda^*)}
\end{equation}
In practice, this can be computed by introducing replicas to treat the
logarithm ($\log x=\lim_{n\to0}\frac\partial{\partial n}x^n$) and introducing another set of replicas
to treat each of the normalizations in the numerator
($x^{-1}=\lim_{m\to-1}x^m$). This leads to the expression
\begin{equation} \label{eq:min.complexity.expanded}
  \begin{aligned}
    \Sigma_{\lambda^*}(E,\mu)
    &=\lim_{\beta\to\infty}\lim_{n\to0}\frac1N\frac\partial{\partial n}
    \int\prod_{a=1}^n\overline{\Bigg[d\nu_H(\mathbf x_a,\pmb\omega_a\mid E,\mu)\,\delta\big(N\lambda^*-(\mathbf s_a^1)^T\operatorname{Hess}H(\mathbf x_a,\pmb\omega_a)\mathbf s_a^1\big)}\\
    &\hspace{12em}\overline{\times\lim_{m_a\to0}
      \left(\prod_{\alpha=1}^{m_a} d\mathbf s_a^\alpha
      \,\delta\big(N-\|\mathbf s_a^\alpha\|^2\big)
      \,\delta\big((\mathbf s_a^\alpha)^T\partial\mathbf g(\mathbf x_a)\big)
      \,e^{-\beta(\mathbf s_a^\alpha)^T\operatorname{Hess}H(\mathbf x_a,\pmb\omega_a)\mathbf s_a^\alpha}\right)
    \Bigg]}
  \end{aligned}
\end{equation}
\end{widetext}
for the complexity of stationary points of a given energy, trace, and smallest eigenvalue.

The marginal complexity follows from the complexity as a function of $\mu$ and
$\lambda^*$ in an analogous way to Section \ref{sec:pseudogap}. In general, one
sets $\lambda^*=0$ and tunes $\mu$ from a sufficiently large value until the
complexity develops an imaginary component, which corresponds to the bulk of
the spectrum touching zero. The value $\mu=\mu_\mathrm m$ that satisfies this
is the marginal stability.

In the cases studied here with zero signal-to-noise, a simpler approach is
possible. The marginal stability $\mu=\mu_\text{m}$ can be identified by
requiring that the complexity is stationary with respect to changes in the
value of the minimum eigenvalue $\lambda^*$, or
\begin{equation} \label{eq:marginal.stability}
  0=\frac\partial{\partial\lambda^*}\Sigma_{\lambda^*}(E,\mu_\text{m}(E))\bigg|_{\lambda^*=0}
\end{equation}
The marginal complexity follows by evaluating the complexity conditioned on
$\lambda^*=0$ at the marginal stability $\mu=\mu_\text{m}(E)$,
\begin{equation} \label{eq:marginal.complexity}
  \Sigma_\text{m}(E)
  =\Sigma_0(E,\mu_\text m(E))
\end{equation}

\subsection{General features of saddle point computation}
\label{sec:general.features}

Several elements of the computation of the marginal complexity, and indeed the
ordinary dominant complexity, follow from the formulae of the above section in
the same way. The physicists' approach to this problem seeks to convert all of
the components of the Kac--Rice measure defined in \eqref{eq:kac-rice.measure} and
\eqref{eq:kac-rice.measure.2} into elements of an exponential integral over
configuration space. To begin with, all Dirac $\delta$ functions are
expressed using their Fourier representation, with
\begin{align}
  \label{eq:delta.grad}
  &\delta\big(\nabla H(\mathbf x_a,\pmb\omega_a)\big)
    =\int\frac{d\hat{\mathbf x}_a}{(2\pi)^N}e^{i\hat{\mathbf x}_a^T\nabla H(\mathbf x_a,\pmb\omega_a)} \\
    \label{eq:delta.energy}
  &\delta\big(NE-H(\mathbf x_a)\big)
    =\int\frac{d\hat\beta_a}{2\pi}e^{\hat\beta_a(NE-H(\mathbf x_a))} \\
  \label{eq:delta.eigen}
  &\begin{aligned}
    &\delta\big(N\lambda^*-(\mathbf s_a^1)^T\operatorname{Hess}H(\mathbf x_a,\pmb\omega)\mathbf s_a^1\big) \\
    &\qquad\qquad\qquad=\int\frac{d\hat\lambda_a}{2\pi}e^{\hat\lambda_a(N\lambda^*-(\mathbf s_a^1)^T\operatorname{Hess}H(\mathbf x_a,\pmb\omega)\mathbf s_a^1)}
  \end{aligned}
\end{align}
To do this we have introduced auxiliary fields $\hat{\mathbf x}_a$,
$\hat\beta_a$, and $\hat\lambda_a$. Because the permutation symmetry of replica vectors
is preserved in \textsc{rsb} orders, the order parameters $\hat\beta$
and $\hat\lambda$ will quickly lose their indices, since they will ubiquitously
be constant over the replica index at the eventual saddle point solution.

We would like to make a similar treatment of the determinant of the Hessian
that appears in \eqref{eq:kac-rice.measure}. The standard approach is to drop
the absolute value function around the determinant. This can potentially lead
to severe problems with the complexity \cite{Fyodorov_2004_Complexity}. However, it is a justified step when
the parameters of the problem $E$, $\mu$, and $\lambda^*$ put us in a
regime where the exponential majority of stationary points have the same index.
This is true for maxima and minima, and for saddle points whose spectra have a
strictly positive bulk with a fixed number of negative outliers. It is in
particular a safe operation for the present problem of marginal minima, which lie
right at the edge of disaster.

Dropping the absolute value function allows us to write
\begin{equation} \label{eq:determinant}
  \det\operatorname{Hess}H(\mathbf x_a, \pmb\omega_a)
  =\int d\bar{\pmb\eta}_a\,d\pmb\eta_a\,
  e^{-\bar{\pmb\eta}_a^T\operatorname{Hess}H(\mathbf x_a,\pmb\omega_a)\pmb\eta_a}
\end{equation}
using $N$-dimensional Grassmann vectors $\bar{\pmb\eta}_a$ and $\pmb\eta_a$. For
the spherical models this step is unnecessary, since there are other ways to
treat the determinant keeping the absolute value signs, as in previous works
\cite{Folena_2020_Rethinking, Kent-Dobias_2023_How}. However, other of
our examples are for models where the same techniques are impossible.

Finally, the $\delta$-function fixing the trace of the Hessian to $\mu$ in
\eqref{eq:kac-rice.measure.2} must be addressed. One could treat it using a
Fourier representation as in (\ref{eq:delta.grad}--\ref{eq:delta.eigen}), but
this is inconvenient because a term of the form
$\operatorname{Tr}\partial\partial H(\mathbf x)$ in the exponential integrand
cannot be neatly captured in superspace representation introduced in the next
section. However, in the cases we study in this paper a simplification can be made: the trace of $\partial\partial H$ can be separated into two pieces, one
that is spatially independent and one that is typically small, or
\begin{equation} \label{eq:mu.star}
  \operatorname{Tr}\partial\partial H(\mathbf x)=N\mu^*_H+\Delta_H(\mathbf x)
\end{equation}
where $\overline{\mu^*_H}=\mu^*$ and $\overline{\Delta_H(\mathbf x)}=O(N^0)$.
Then fixing the trace of the Hessian to $\mu$ implies that
\begin{equation}
  \begin{aligned}
    \mu
    &=\frac1N\operatorname{Tr}\operatorname{Hess}H(\mathbf x)
    =\frac1N\left(\partial\partial H(\mathbf x)+
    \sum_{i=1}^r\omega_i\operatorname{Tr}\partial\partial g_i(\mathbf x)\right)
    \\
    &=\mu^*+\frac1N\sum_{i=1}^r\omega_i\operatorname{Tr}\partial\partial g_i(\mathbf x)
    +O(N^{-1})
  \end{aligned}
\end{equation}
for typical samples $H$.
In particular, here we study only cases with quadratic $g_i$, which results in
a linear expression relating $\mu$ and the $\omega_i$ that is independent of $\mathbf
x$. Since $H$ contains the disorder of the problem, this simplification means
that the effect of fixing the trace is largely independent of the disorder and mostly
depends on properties of the constraint manifold.

\subsection{Superspace representation}
\label{sec:superspace_kac-rice}

The ordinary Kac--Rice calculation involves many moving parts, and this method
for incorporating marginality adds even more. It is therefore convenient to
introduce compact and simplifying notation through a superspace representation.
The use of superspace in the Kac--Rice calculation is well established, as well
as the deep connections with BRST symmetry that is implied \cite{Annibale_2003_Supersymmetric, Annibale_2003_The, Annibale_2004_Coexistence}.
Appendix~\ref{sec:superspace} introduces the notation and methods of
superspace algebra. Here we describe how it can be used to simplify the complexity
calculation for marginal minima.

We consider the $\mathbb R^{N|4}$ superspace whose Grassmann indices are
$\bar\theta_1,\theta_1,\bar\theta_2,\theta_2$. Consider the supervector defined
by
\begin{equation}
  \pmb\phi_a^\alpha(1,2)
  =\mathbf x_a
  +\bar\theta_1\pmb\eta_a+\bar{\pmb\eta}_a\theta_1
  +i\hat{\mathbf x}_a\bar\theta_1\theta_1
  +\mathbf s_a^\alpha(\bar\theta_1\theta_2+\bar\theta_2\theta_1)
\end{equation}
Note that this supervector does not span the whole superspace: only a couple
terms from the $\bar\theta_2,\theta_2$ sector are present, since the rest are
unnecessary for our representation. With this supervector so defined, the
replicated count of stationary points with energy $E$, trace $\mu$, and
smallest eigenvalue $\lambda^*$ can be written as
\begin{widetext}
\begin{equation}
  \begin{aligned}
    \mathcal N_H(E,\mu,\lambda^*)^n
    &=\lim_{\beta\to\infty}\int d\pmb\omega\,d\hat\beta\,d\hat\lambda\prod_{a=1}^n\lim_{m_a\to0}\prod_{\alpha=1}^{m_a}d\pmb\phi_a^\alpha
    \exp\left\{
      \delta^{\alpha1}N(\hat\beta E+\hat\lambda\lambda^*)
      +\int d1\,d2\,B^\alpha(1,2)L\big(\pmb\phi_a^\alpha(1,2),\pmb\omega\big)
    \right\}
  \end{aligned}
\end{equation}
Here we have also defined the operator
\begin{equation}
  B^\alpha(1,2)=\delta^{\alpha1}\bar\theta_2\theta_2
        (1-\hat\beta\bar\theta_1\theta_1)
        -\delta^{\alpha1}\hat\lambda-\beta
\end{equation}
which encodes various aspects of the complexity problem. When the Lagrangian is
expanded in a series with respect to the Grassmann indices and the definition
of $B$ inserted, the result of the Grassmann integrals produces exactly the
content of the integrand in \eqref{eq:min.complexity.expanded} with the
substitutions \eqref{eq:delta.grad}, \eqref{eq:delta.energy},
\eqref{eq:delta.eigen}, and \eqref{eq:determinant} of the Dirac $\delta$
functions and the determinant made. The new measures
\begin{align}
  d\pmb\phi_a^\alpha
  &=\left[
    d\mathbf x_a\,\delta\big(\mathbf g(\mathbf x_a)\big)\,
    \frac{d\hat{\mathbf x}_a}{(2\pi)^N}\,
    d\pmb\eta_a\,d\bar{\pmb\eta}_a\,
    \delta^{\alpha1}+(1-\delta^{\alpha1})
  \right]\,
  d\mathbf s_a^\alpha\,\delta(\|\mathbf s_a^\alpha\|^2-N)\,
  \delta\big((\mathbf s_a^\alpha)^T\partial\mathbf g(\mathbf x_a)\big)
  \\
  d\pmb\omega&=\bigg(\prod_{i=1}^rd\omega_i\bigg)
  \,\delta\bigg(N\mu-\mu^*-\sum_i^r\omega_i\operatorname{Tr}\partial\partial g_i\bigg)
\end{align}
collect the individual measures of the various fields embedded in the superfield, along with their constraints.
\end{widetext}
With this way of writing the replicated count, the problem of marginal
complexity temporarily takes the schematic form of an equilibrium calculation
with configurations $\pmb\phi$, inverse temperature $B$, and energy $L$. This
makes the intermediate pieces of the calculation dramatically simpler. Of
course the intricacies of the underlying problem are not banished: near the end
of the calculation, terms involving the superspace must be expanded.
We will make use of this representation to simplify the analysis of the marginal complexity when analyzing random sums of squares in Section \ref{sec:least.squares}.

\section{Examples}
\label{sec:examples}

In this section we present analysis of marginal complexity in three random
landscapes. In Section \ref{sec:ex.spherical} we treat the spherical spin glasses, which reveals some general aspects of the
calculation. Since the spherical spin glasses are Gaussian and have identical
GOE spectra at each stationary point, the approach introduced here is overkill.
In Section \ref{sec:multispherical} we apply the methods to a multispherical
spin glass, which is still Gaussian but has a non-GOE spectrum whose shape can vary
between stationary points. Finally, in Section \ref{sec:least.squares} we analyze a model of sums of squared random functions, which is non-Gaussian and whose Hessian statistics depend on the conditioning of the energy and gradient.

\subsection{Spherical spin glasses}
\label{sec:ex.spherical}

The spherical spin glasses are a family of models that encompass every
isotropic Gaussian field on the hypersphere. Their configuration space is the sphere $S^{N-1}$ defined by all $\mathbf x\in\mathbb
R^N$ such that $0=g(\mathbf x)=\frac12(\|\mathbf x\|^2-N)$. One can consider the models as
defined by ensembles of centered Gaussian functions $H$ such that the covariance between two
points in the configuration space is
\begin{equation}
  \overline{H(\mathbf x)H(\mathbf x')}=Nf\left(\frac{\mathbf x\cdot\mathbf x'}N\right)
\end{equation}
for some function $f$ with positive series coefficients. Such functions can be considered to be made up of all-to-all tensorial interactions, with
\begin{equation}
  H(\mathbf x)
  =\sum_{p=0}^\infty\frac1{p!}\sqrt{\frac{f^{(p)}(0)}{N^{p-1}}}
  \sum_{i_1\cdots i_p}^NJ_{i_1\cdots i_p}x_{i_1}\cdots x_{i_p}
\end{equation}
and the elements of the tensors $J$ being independently distributed with the
unit normal distribution \cite{Crisanti_1993_The}. We focus on marginal minima
in models with $f'(0)=0$, which corresponds to models without a random external
field. Such a random field would correspond in each individual sample $H$ to a
signal, and therefore complicate the analysis by correlating the positions of
stationary points and the eigenvectors of their Hessians. Here, $\mu^*$ of
\eqref{eq:mu.star} is zero.

The marginal optima of these models can be studied without the methods
introduced in this paper, and have been in the past \cite{Folena_2020_Rethinking,
Kent-Dobias_2023_How}. First, these models are Gaussian, so at large $N$ the
Hessian is statistically independent of the gradient and energy
\cite{Fyodorov_2004_Complexity, Bray_2007_Statistics}. Therefore, conditioning the Hessian can be done
mostly independently from the problem of counting stationary points. Second, in
these models the Hessian at every point in the landscape belongs to the GOE
class with the same width of the spectrum $\mu_\mathrm m=2\sqrt{f''(1)}$.
Therefore, all marginal minima in these systems have the same constant shift
$\mu=\mu_\mathrm m$. Despite the fact that the complexity of marginal optima is
well known by simpler methods, it is instructive to carry through the
calculation for this case, since we will learn some things about its application in
more nontrivial settings.

Note that in the pure version of these models with $f(q)=\frac12q^p$, the
methods of this section must be amended slightly. This is because in these
models there is an exact correspondence $\mu=-pE$ between the trace of the
Hessian and the energy, and therefore they cannot be fixed independently. This
correspondence implies that when $\mu=\mu_\mathrm m$, the corresponding energy level
$E_\mathrm{th}=-\frac1p\mu_\mathrm m$ contains all marginal minima. This is what gives this threshold energy such singular importance to dynamics in the pure spherical models.

The procedure to treat the complexity of the spherical models has been made in
detail elsewhere \cite{Kent-Dobias_2023_How}. Here we make only a sketch of the
steps involved. First we notice that
$\mu=\frac1N\omega\operatorname{Tr}\partial\partial g(\mathbf x)=\omega$, so
that the only Lagrange multiplier $\omega$ in this problem is set directly to
the shift $\mu$. The substitutions \eqref{eq:delta.grad},
\eqref{eq:delta.energy}, and \eqref{eq:delta.eigen} are made to convert the
Dirac $\delta$ functions into exponential integrals, and the substitution
\eqref{eq:determinant} is made to likewise convert the determinant.

Once these substitutions have been made, the entire expression
\eqref{eq:min.complexity.expanded} is an exponential integral whose argument is
a linear functional of $H$. This allows for the average to be taken over the
disorder. If we gather all the $H$-dependant pieces associated with replica $a$
into the linear functional $\mathcal O_a$ then the average over the ensemble of functions $H$ gives
\begin{equation}
  \begin{aligned}
    \overline{
      e^{\sum_a^n\mathcal O_aH(\mathbf x_a)}
    }
    &=e^{\frac12\sum_a^n\sum_b^n\mathcal O_a\mathcal O_b\overline{H(\mathbf x_a)H(\mathbf x_b)}} \\
    &=e^{N\frac12\sum_a^n\sum_b^n\mathcal O_a\mathcal O_bf\big(\frac{\mathbf x_a\cdot\mathbf x_b}N\big)}
  \end{aligned}
\end{equation}
The result is an integrand that depends on the many vector variables we
have introduced only through their scalar products with each other. We therefore make a change of variables in the integration from those vectors to matrices that encode their possible scalar products. These matrices are
\begin{align}
    &C_{ab}=\frac1N\mathbf x_a\cdot\mathbf x_b
    \quad
    &R_{ab}=-i\frac1N\mathbf x_a\cdot\hat{\mathbf x}_b
    \quad
    &D_{ab}=\frac1N\hat{\mathbf x}_a\cdot\hat{\mathbf x}_b& \notag \\
    &Q_{ab}^{\alpha\gamma}=\frac1N\mathbf s_a^\alpha\cdot\mathbf s_b^\gamma
    \quad
    &\hat X^\alpha_{ab}=-i\frac1N\hat{\mathbf x}_a\cdot\mathbf s_b^\alpha
    \quad
    &X^\alpha_{ab}=\frac1N\mathbf x_a\cdot\mathbf s_b^\alpha&
    \notag \\
    &G_{ab}=\frac1N\bar{\pmb\eta}_a\cdot\pmb\eta_b
    \label{eq:order.parameters}
\end{align}
Order parameters that mix the normal and Grassmann variables generically vanish
in these settings and we don't consider them here \cite{Kurchan_1992_Supersymmetry}.
This transformation changes the measure of the integral, with
\begin{equation}
  \begin{aligned}
    &\prod_{a=1}^nd\mathbf x_a\,\frac{d\hat{\mathbf x}_a}{(2\pi)^N}\,
    d\bar{\pmb\eta}_a\,d\pmb\eta_a\,\prod_{\alpha=1}^{m_a}d\mathbf s_a^\alpha \\
    &\quad=dC\,dR\,dD\,dG\,dQ\,dX\,d\hat X\,(\det J)^{N/2}(\det G)^{-N}
  \end{aligned}
\end{equation}
where $J$ is the Jacobian of the transformation in the real-valued fields. This
Jacobian takes a block form
\begin{equation} \label{eq:coordinate.jacobian}
  J=\begin{bmatrix}
    C&iR&X_1&\cdots&X_n \\
    iR&D&i\hat X_1&\cdots&i\hat X_n\\
    X_1^T&i\hat X_1^T&Q_{11}&\cdots&Q_{1n}\\
    \vdots&\vdots&\vdots&\ddots&\vdots\\
    X_n^T&i\hat X_n^T&Q_{n1}&\cdots&Q_{nn}
  \end{bmatrix}
\end{equation}
The Grassmann integrals produces their own inverted
Jacobian. The matrix that make up the blocks of the matrix $J$ are such that $C$, $R$, and $D$ are $n\times n$ matrices indexed by their lower indices, $Q_{ab}$ is an
$m_a\times m_b$ matrix indexed by its upper indices, while $X_a$ is an $n\times
m_a$ matrix with one lower and one upper index.

These steps follow identically to those more carefully outlined in
the cited papers \cite{Folena_2020_Rethinking, Kent-Dobias_2023_How}. Following them in the present case, we arrive
at a form for the complexity of stationary points with fixed energy $E$, stability $\mu$, and lowest eigenvalue $\lambda^*$ with
\begin{widetext}
  \begin{equation} \label{eq:spherical.complexity}
  \begin{aligned}
    &\Sigma_{\lambda^*}(E,\mu)
    =\lim_{\beta\to\infty}\lim_{n\to0}\lim_{m_1\cdots m_n\to0}
    \frac1N\frac\partial{\partial n}
    \int dC\,dR\,dD\,dG\,dQ\,dX\,d\hat X\,d\hat\beta\,d\hat\lambda\,
    \exp\Bigg\{
      nN\mathcal S_\mathrm{SSG}(\hat\beta,C,R,D,G\mid E,\mu) \\
      &\qquad
      +nN\mathcal U_\mathrm{SSG}(\hat\lambda,Q,X,\hat X\mid\beta,\lambda^*,\mu,C)
      +\frac N2\log\det\left[
        I-\begin{bmatrix}
          Q_{11}&\cdots&Q_{1n}\\
          \vdots&\ddots&\vdots\\
          Q_{n1}&\cdots&Q_{nn}
        \end{bmatrix}^{-1}
        \begin{bmatrix}
          X_1^T&i\hat X_1^T\\
          \vdots&\vdots\\
          X_n^T&i\hat X_n^T
        \end{bmatrix}
        \begin{bmatrix}
          C&iR\\iR&D
        \end{bmatrix}^{-1}
        \begin{bmatrix}
          X_1\cdots X_n\\
          i\hat X_1\cdots i\hat X_n
        \end{bmatrix}
      \right]
    \Bigg\}
  \end{aligned}
\end{equation}
The exponential integrand is split into two effective actions coupled only by a
residual determinant. The first of these actions is the usual effective action
for the complexity of the spherical spin glasses, or
\begin{equation} \label{eq:spherical.action}
  \begin{aligned}
    &\mathcal S_\mathrm{SSG}(\hat\beta,C,R,D,G\mid E,\mu)
    =\hat\beta E+\lim_{n\to0}\frac1n\bigg\{-\mu\operatorname{Tr}(R+G) \\
    &\qquad+\frac12\sum_{ab}\left(
      \hat\beta^2f(C_{ab})
      +\big(2\hat\beta R_{ab}-D_{ab}\big)f'(C_{ab})
      +(R_{ab}^2-G_{ab}^2)f''(C_{ab})
    \right)
    +\frac12\log\det\begin{bmatrix}C&iR\\iR^T&D\end{bmatrix}
    -\log\det G\bigg\}
  \end{aligned}
\end{equation}
The second of these actions is analogous to the effective action \eqref{eq:goe.action} from the GOE example of Section~\ref{sec:shifted.GOE} and contains
the contributions from the marginal pieces of the calculation, and is given by
\begin{equation}
  \begin{aligned}
    &\mathcal U_\mathrm{SSG}(\hat\lambda,Q,X,\hat X\mid\beta,\lambda^*,\mu,C)
    =\hat\lambda\lambda^*
    +\lim_{n\to0}\lim_{m_1\cdots m_n\to0}\frac1n\Bigg\{
    \frac12\log\det Q-
      \sum_{a=1}^n\bigg(
        \sum_{\alpha=1}^{m_a}\beta\mu Q_{aa}^{\alpha\alpha}
        +\hat\lambda\mu Q_{aa}^{11}
      \bigg)
    +2\sum_{ab}^nf''(C_{ab})
    \\
     &\quad\times\Bigg[\beta\sum_\alpha^{m_a}\left(
        \beta\sum_\gamma^{m_b}(Q_{ab}^{\alpha\gamma})^2
        -\hat\beta(X_{ab}^\alpha)^2
        -2X_{ab}^\alpha\hat X_{ab}^\alpha
      \right)
    +\hat\lambda\left(
      \hat\lambda(Q_{ab}^{11})^2
      -\hat\beta(X_{ab}^1)^2
      -2X_{ab}^1\hat X_{ab}^1
    \right)
    +\beta\hat\lambda\left(
        \sum_\alpha^{m_a} Q_{ab}^{\alpha1}
        +\sum_\alpha^{m_b} Q_{ab}^{1\alpha}
    \right)\Bigg]
    \Bigg\}
  \end{aligned}
\end{equation}
\end{widetext}
The fact that the complexity can be split into two relatively independent
pieces in this way is a characteristic of the isotropic and Gaussian nature of the spherical
spin glass. In Section \ref{sec:least.squares} we will study a model whose
energy is isotropic but not Gaussian and where such a decomposition is impossible.

There are some dramatic simplifications that emerge from the structure of this
particular problem. First, notice that the dependence on the parameters $X$ and $\hat X$ are purely quadratic.
Therefore, there will always be a saddle point condition where they are both
zero. In this case without a fixed or random field, we except this solution to be correct. We can reason about
why this is so: $X$, for instance, quantifies the correlation between the
typical position of stationary points and the direction of their typical
eigenvectors. In a landscape without a signal, where no direction is any more
important than any other, we expect such correlations to be zero:
where a state is located does not give any information as to the orientation
of its soft directions. On the other hand, in the spiked case, or with an
external field, the preferred direction can polarize both the direction of
typical stationary points \emph{and} their soft eigenvectors. Therefore, in
these instances one must account for solutions with nonzero $X$ and $\hat X$.

We similarly expect that $Q_{ab}=0$ for $a\neq b$. For the contrary to be true,
eigenvectors at independently sampled stationary points would need to have
their directions correlated. This is expected in situations with a signal,
where such correlations would be driven by a shared directional bias towards
the signal. In the present situation, where there is no signal, such
correlations do not exist.

When we take $X=\hat X=0$ and
$Q^{\alpha\beta}_{ab}=\delta_{ab}Q^{\alpha\beta}$, we find that
\begin{equation}
  \mathcal U_\mathrm{SSG}(\hat\lambda,Q,0,0\mid\beta,\lambda^*,\mu,C)
  =\mathcal U_\mathrm{GOE}(\hat\lambda,Q\mid\beta,\lambda^*,\mu)
\end{equation}
with $\sigma^2=f''(1)$. That is, the effective action for the terms related to
fixing the eigenvalue in the spherical Kac--Rice problem is exactly the same as
that for the \textrm{GOE} problem. This is perhaps not so surprising, since we
established from the beginning that the Hessian of the spherical spin glasses
belongs to the GOE class.

The remaining analysis of the eigenvalue-dependent part $\mathcal
U_\mathrm{SSG}$ follows precisely the same steps as were made in Section
\ref{sec:shifted.GOE} for the GOE example. The result of the calculation is
also the same: the exponential factor containing $\mathcal U_\mathrm{SSG}$
produces precisely the large deviation function $G_{\lambda^*}(\mu)$ of
\eqref{eq:goe.large.dev} (again with $\sigma^2=f''(1)$). The remainder of the
integrand depending on $\mathcal S_\mathrm{SSG}$ produces the ordinary
complexity of the spherical spin glasses without conditions on the Hessian
eigenvalue. We therefore find that
\begin{equation}
  \Sigma_{\lambda^*}(E,\mu)
  =\Sigma(E,\mu)+G_{\lambda^*}(\mu)
\end{equation}
We find the marginal complexity by solving
\begin{equation}
  0
  =\frac\partial{\partial\lambda^*}\Sigma_{\lambda^*}(E,\mu_\mathrm m(E))\bigg|_{\lambda^*=0}
  =\frac\partial{\partial\lambda^*}G_{\lambda^*}(\mu_\mathrm m(E))\bigg|_{\lambda^*=0}
\end{equation}
which gives $\mu_\mathrm m(E)=2\sigma=2\sqrt{f''(1)}$ independent of $E$, as we
presaged above. Since $G_0(\mu_\mathrm m)=0$, this gives finally
\begin{equation}
  \Sigma_\mathrm m(E)
  =\Sigma_0(E,\mu_\mathrm m(E))
  =\Sigma(E,\mu_\mathrm m)
\end{equation}
that the marginal complexity in these models is simply the ordinary complexity
evaluated at a fixed trace $\mu_\mathrm m$ of the Hessian.

\subsection{Multispherical spin glasses}
\label{sec:multispherical}

The multispherical spin glasses are a simple extension of the spherical ones, where
the configuration space is taken to be the union of more than one hypersphere.
Here we consider the specific case where the configuration space is the union
of two $(N-1)$-spheres, with $\Omega=S^{N-1}\times S^{N-1}$. The two spheres
give rise to two constraints: for $\mathbf x=[\mathbf x^{(1)},\mathbf x^{(2)}]$
with components $\mathbf x^{(1)},\mathbf x^{(2)}\in\mathbb R^N$, the
constraints are $0=g_1(\mathbf x)=\frac12(\|\mathbf x^{(1)}\|^2-N)$ and
$0=g_2(\mathbf x)=\frac12(\|\mathbf x^{(2)}\|^2-N)$. These two constraints are
fixed by two Lagrange multipliers $\omega_1$ and $\omega_2$.

The energy in our multispherical spin glass is given by
\begin{equation}
  H(\mathbf x)=H_1(\mathbf x^{(1)})+H_2(\mathbf x^{(2)})-\epsilon\mathbf x^{(1)}\cdot\mathbf x^{(2)}
\end{equation}
The energy $H_i$ of each individual sphere is taken to be a centered Gaussian
random function with a covariance given in the usual spherical spin glass way
for $\mathbf x,\mathbf x'\in\mathbb R^N$ by
\begin{equation}
  \overline{H_i(\mathbf x)H_j(\mathbf x')}
  =N\delta_{ij}f_i\left(\frac{\mathbf x\cdot\mathbf x'}N\right)
\end{equation}
with the functions $f_1$ and $f_2$ not necessarily the same. As for the spherical spin glasses, $\mu^*$ of \eqref{eq:mu.star} is zero.

In this problem, there is an energetic competition between the independent spin
glass energies on each sphere and their tendency to align or anti-align through
the interaction term. These models have more often been studied with random
fully connected couplings between the spheres, for which it is possible to also
use configuration spaces involving spheres of different sizes
\cite{Subag_2021_TAP, Subag_2023_TAP, Bates_2022_Crisanti-Sommers,
  Bates_2022_Free, Huang_2023_Strong, Huang_2023_Algorithmic,
Huang_2024_Optimization}. As far as we are aware, the deterministically coupled model has not been previously studied, except as a thought experiment in \cite{Kent-Dobias_2023_How}.

We again make use of the method of Lagrange multipliers to find stationary points on the constrained configuration space. The Lagrangian and its gradient and Hessian are
\begin{align}
  &\begin{aligned}
    L(\mathbf x)&=H(\mathbf x)
    +\frac12\omega_1\big(\|\mathbf x^{(1)}\|^2-N\big) \\
                &\qquad\qquad\qquad+\frac12\omega_2\big(\|\mathbf x^{(2)}\|^2-N\big)
  \end{aligned}
  \\
  &\nabla H(\mathbf x,\pmb\omega)
              =
              \begin{bmatrix}
                \partial_1H_1(\mathbf x^{(1)})-\epsilon\mathbf x^{(2)}+\omega_1\mathbf x^{(1)} \\
                \partial_2H_2(\mathbf x^{(2)})-\epsilon\mathbf x^{(1)}+\omega_2\mathbf x^{(2)}
  \end{bmatrix}
  \\
  &\begin{aligned}
              &\operatorname{Hess}H(\mathbf x,\pmb\omega) \\
              &\quad=
              \begin{bmatrix}
  \partial_1\partial_1H_1(\mathbf x^{(1)})+\omega_1I&-\epsilon I \\
  -\epsilon I&\partial_2\partial_2H_2(\mathbf x^{(2)})+\omega_2I
  \end{bmatrix}
  \end{aligned}
\end{align}
where $\partial_1=\frac\partial{\partial\mathbf x^{(1)}}$ and $\partial_2=\frac\partial{\partial\mathbf x^{(2)}}$.
Like in the spherical spin glasses, fixing the trace of the Hessian to $\mu$ is
equivalent to a constraint on the Lagrange multipliers. However, in this case
it corresponds to $\mu=\omega_1+\omega_2$, and therefore they are not
uniquely fixed by fixing $\mu$.

Since the energy in the multispherical models is Gaussian, the properties of
the matrix $\partial\partial H$ are again independent of the energy and
gradient. This means that the form of the Hessian is parameterized solely by
the values of the Lagrange multipliers $\omega_1$ and $\omega_2$, just
as $\mu=\omega$ alone parameterized the Hessian in the spherical spin glasses.
Unlike that case, however, the Hessian takes different shapes with different
spectral widths depending on their precise combination. In
Appendix~\ref{sec:multispherical.spectrum} we derive a variational form for the
spectral density of the Hessian in these models using standard methods.

Because of the independence of the Hessian, the method introduced in this
article is not necessary to characterize the marginal minima of this system.
Rather, we could take the spectral density derived in
Appendix~\ref{sec:multispherical.spectrum} and find the Lagrange multipliers
$\omega_1$ and $\omega_2$ corresponding with marginality by tuning the edge of
the spectrum to zero. In some ways the current method is more convenient than
this, since it is a purely variational method and therefore can be reduced to a
single root-finding exercise.

Unlike the constraints on the configurations $\mathbf x$, the constraint on the
tangent vectors $\mathbf s=[\mathbf s^{(1)},\mathbf s^{(2)}]\in\mathbb R^{2N}$
remains the same spherical constraint as before, which implies $N=\|\mathbf
s\|^2=\|\mathbf s^{(1)}\|^2+\|\mathbf s^{(2)}\|^2$. Defining intra- and inter-sphere overlap matrices
\begin{equation}
  Q^{ij,\alpha\gamma}_{ab}=\frac1N\mathbf s^{(i),\alpha}_a\cdot\mathbf s^{(j),\gamma}_b
\end{equation}
this problem no longer has the property that the diagonal of the $Q$s is one,
but instead that $1=Q^{11,\alpha\alpha}_{aa}+Q^{22,\alpha\alpha}_{aa}$. This is
the manifestation of the fact that a normalized vector in the tangent space of the
multispherical model need not be equally spread on the two subspaces, but can
be concentrated in one or the other.

The calculation of the marginal complexity in this problem follows very closely
to that of the spherical spin glasses in the previous subsection. We
immediately make the simplifying assumptions that the soft directions of different
stationary points are typically uncorrelated and therefore $X=\hat X=0$ and the
overlaps $Q$ between eigenvectors are only nonzero when in the same replica.
The result for the complexity has the schematic form of \eqref{eq:spherical.complexity}, but with
different effective actions depending now on overlaps inside each of the two
spheres and between the two spheres. The effective action for the traditional
complexity of the multispherical spin glass is
\begin{widetext}
\begin{align}
    &\mathcal S_\mathrm{MSG}(\hat\beta,C^{11},R^{11},D^{11},G^{11},C^{22},R^{22},D^{22},G^{22},C^{12},R^{12},R^{21},D^{12},G^{12},G^{21}
    \mid E,\omega_1,\omega_2)= \hat\beta(E-E_1-E_2-\epsilon c_d^{12}) \notag \\
    &
    +\mathcal S_\mathrm{SSG}(\hat\beta,C^{11},R^{11},D^{11},G^{11}\mid E_1,\omega_1)
    +\mathcal S_\mathrm{SSG}(\hat\beta,C^{22},R^{22},D^{22},G^{22}\mid E_2,\omega_2)
    +\lim_{n\to0}\frac1n\Bigg\{
      \epsilon\operatorname{Tr}(R^{12}+R^{21}+G^{12}+G^{21}-\hat\beta C^{12})
    \notag \\
    &\quad
      +\frac12\log\det\left(
    I-
    \begin{bmatrix}C^{11}&iR^{11}\\iR^{11}&D^{11}\end{bmatrix}^{-1}
    \begin{bmatrix}
      C^{12} & iR^{12} \\ iR^{21} & D^{12}
    \end{bmatrix}
    \begin{bmatrix}C^{22}&iR^{22}\\iR^{22}&D^{22}\end{bmatrix}^{-1}
    \begin{bmatrix}
      C^{12} & iR^{21} \\ iR^{21} & D^{12}
    \end{bmatrix}
    \right)
    -\log\det(I-(G^{11}G^{22})^{-1}G^{12}G^{21})\Bigg\}
\end{align}
which is the sum of two effective actions \eqref{eq:spherical.action} for the spherical spin glass
associated with each individual sphere, and some coupling terms. The order
parameters are defined the same as in the spherical spin glasses, but now with
raised indices to indicate whether the vectors come from one or the other
spherical subspace. The effective action for the eigenvalue-dependent part of
the complexity is likewise given by
\begin{align}
    &\mathcal U_\mathrm{MSG}(\hat q,\hat\lambda,Q^{11},Q^{22},Q^{12}\mid\beta,\lambda^*,\omega_1,\omega_2) \notag \\
    &\quad=\lim_{m\to0}\bigg\{\sum_{\alpha=1}^m\left[\hat q^\alpha(Q^{11,\alpha\alpha}+Q^{22,\alpha\alpha}-1)-\beta(\omega_1Q^{11,\alpha\alpha}+\omega_2Q^{22,\alpha\alpha}-2\epsilon Q^{12,\alpha\alpha})\right]
    -\hat\lambda(\omega_1Q^{11,11}+\omega_2Q^{22,11}-2\epsilon Q^{12,11}) \notag \\
    &\qquad\qquad+\sum_{i=1,2}f_i''(1)\left[\beta^2\sum_{\alpha\gamma}^m(Q^{ii,\alpha\gamma})^2+2\beta\hat\lambda\sum_\alpha^m(Q^{ii,1\alpha})^2+\hat\lambda^2(Q^{ii,11})^2\right]
    +\frac12\log\det\begin{bmatrix}
      Q^{11}&Q^{12}\\
      Q^{12}&Q^{22}
    \end{bmatrix}
  \bigg\}
\label{eq:multispherical.marginal.action}
\end{align}
\end{widetext}
The new variables $\hat q^\alpha$ are Lagrange multipliers introduced
to enforce the constraint that $Q^{11,\alpha\alpha}+Q^{22,\alpha\alpha}=1$.
Because of this constraint, the diagonal of the $Q$ matrices cannot be
taken to be 1 as in Section~\ref{sec:shifted.GOE}. Instead we take each of the matrices $Q^{11}$, $Q^{22}$, and $Q^{12}$ to have
the planted replica symmetric form of \eqref{eq:Q.structure}, but with the
diagonal not necessarily equal to 1, so
\begin{equation}
 Q^{ij}=\begin{bmatrix}
   \tilde q^{ij}_d & \tilde q^{ij}_0 & \tilde q^{ij}_0 & \cdots & \tilde q^{ij}_0 \\
   \tilde q^{ij}_0 & q^{ij}_d & q^{ij}_0 & \cdots & q^{ij}_0 \\
   \tilde q^{ij}_0 & q^{ij}_0 & q^{ij}_d & \ddots & q^{ij}_0 \\
   \vdots & \vdots & \ddots & \ddots & \vdots \\
   \tilde q^{ij}_0 & q^{ij}_0 & q^{ij}_0 & \cdots & q^{ij}_d
 \end{bmatrix}
\end{equation}
This requires us to introduce two new order parameters $\tilde q^{ij}_d$ and
$q^{ij}_d$ per pair $(i,j)$, in addition to the off-diagonal order parameters
$\tilde q_0^{ij}$ and $q_0^{ij}$ already present in \eqref{eq:Q.structure}. We
also need two separate Lagrange multipliers $\hat q$ and $\hat{\tilde q}$ to
enforce the tangent space normalization $q_d^{11}+q_d^{22}=1$ and $\tilde q_d^{11}+\tilde q_d^{22}=1$ for the tilde and untilde replicas,
respectively, which will in general take different values at the saddle point.
When
this ansatz is inserted into the expression \eqref{eq:multispherical.marginal.action} for the effective action and the
limit of $m\to0$ is taken, we find
\begin{widetext}
  \begin{align}
      &\mathcal U_\mathrm{MSG}(\hat q,\hat{\tilde q},\hat\lambda,\tilde q_d^{11},\tilde q_0^{11},q_d^{11},q_0^{11},\tilde q_d^{22},\tilde q_0^{22},q_d^{22},q_0^{22},\tilde q_d^{12},\tilde q_0^{12},q_d^{12},q_0^{12}\mid\beta,\lambda^*,\omega_1,\omega_2) \notag \\
      &=\sum_{i=1,2}\left\{f_i''(1)\left[
        \beta^2\left(
          (\tilde q^{ii}_d)^2
          -(q^{ii}_d)^2
          +2(q^{ii}_0)^2
          -2(\tilde q^{ii}_0)^2
        \right)
        +2\beta\hat\lambda\left(
          (\tilde q^{ii}_d)^2-(\tilde q^{ii}_0))^2
        \right)
        +\hat\lambda^2(\tilde q^{ii}_d)^2
      \right]
      -\hat\lambda\tilde q^{ii}_d\omega_i
      -\beta(\tilde q^{ii}_d-q^{ii}_d)\omega_i
      \right\} \notag \\
      &+\frac12\log\bigg[
        \left(
          2q^{12}_0\tilde q^{12}_0-\tilde q^{12}_0(\tilde q^{12}_d+q^{12}_d)
          -2\tilde q^{11}_0q^{22}_0+\tilde q^{11}_d\tilde q^{22}_0+\tilde q^{11}_0q^{22}_d
        \right)
        \left(
          2q^{12}_0\tilde q^{12}_0-\tilde q^{12}_0(\tilde q^{12}_d+q^{12}_d)
          -2q^{11}_0\tilde q^{22}_0+q^{11}_d\tilde q^{22}_0+\tilde q^{11}_0\tilde q^{22}_d
        \right) \notag \\
      &\qquad\qquad+2\left(3(q^{12}_0)^2-(\tilde q^{12}_0)^2-2q^{12}_0q^{12}_d-3q^{11}_0q^{22}_0+q^{11}_dq^{22}_0+\tilde q^{11}_0\tilde q^{22}_0+q^{11}_0q^{22}_d
        \right)\left(
          (\tilde q^{12}_0)^2-(\tilde q^{12}_d)^2-\tilde q^{11}_0\tilde q^{22}_0+\tilde q^{11}_d\tilde q^{22}_d
        \right) \notag \\
      &\qquad\qquad-\left(
        2(q^{12}_0)^2-(\tilde q^{12}_0)^2-(q^{12}_d)^2-2q^{11}_0q^{22}_0+\tilde q^{11}_0\tilde q^{22}_0+q^{11}_dq^{22}_d
      \right)\left(
        (\tilde q^{12}_0)^2-(\tilde q^{12}_d)^2-\tilde q^{11}_0\tilde q^{22}_0+\tilde q^{11}_d\tilde q^{22}_d
      \right)
      \bigg]
      \notag \\
      &-\log\left[(q^{11}_d-q^{11}_0)(q^{22}_d-q^{22}_0)-(q^{12}_d-q^{12}_0)^2\right]
      +2\epsilon\big[\hat\lambda\tilde q^{12}_d
        +\beta(\tilde q^{12}_d-q^{12}_d)\big]
      -\hat q(q^{11}_d+q^{22}_d-1)+\hat{\tilde q}(\tilde q^{11}_d+\tilde q^{22}_d-1)
      \label{eq:multispherical.ansatz}
  \end{align}
\end{widetext}
To make the limit to zero temperature, we once again need an ansatz for the
asymptotic behavior of the overlaps. These take the form
$q^{ij}_0=q^{ij}_d-y^{ij}_0\beta^{-1}-z^{ij}_0\beta^{-2}$. Notice that in this case, the asymptotic behavior of the
off-diagonal elements is to approach the value of the diagonal rather than to approach one.
We also require $\tilde q^{ij}_d=q^{ij}_d-\tilde y^{ij}_d\beta^{-1}-\tilde
z^{ij}_d\beta^{-2}$, i.e., that the tilde diagonal terms also approach the
same diagonal value as the untilde terms, but with potentially different rates.

As before, in order for the logarithmic term to stay finite, there are necessary
constraints on the values $y$. These are
\begin{align}
  \frac12(y^{11}_d-\tilde y^{11}_d)=y^{11}_0-\tilde y^{11}_0 \\
  \frac12(y^{22}_d-\tilde y^{22}_d)=y^{22}_0-\tilde y^{22}_0 \\
  \frac12(y^{12}_d-\tilde y^{12}_d)=y^{12}_0-\tilde y^{12}_0
\end{align}
One can see that when the diagonal elements are all equal, this requires the
$y$s for the off-diagonal elements to be equal, as in the GOE case. Here, since
the diagonal elements are not necessarily equal, we have a more general
relationship.

\begin{figure}
  \includegraphics{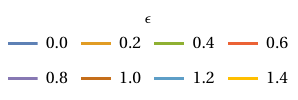}

  \vspace{1em}

  \includegraphics{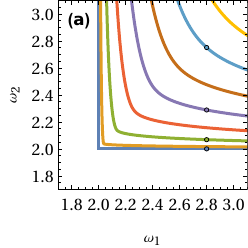}
  \hfill
  \includegraphics{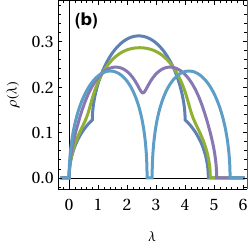}

  \vspace{1em}

  \includegraphics{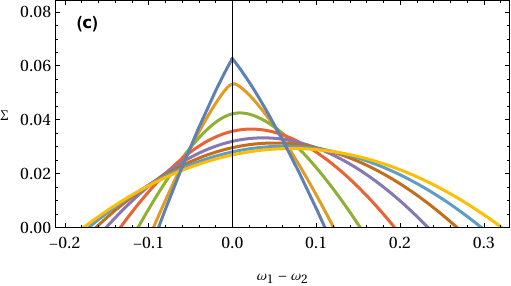}

  \caption{
    Properties of marginal minima in the multispherical model.
    \textbf{(a)}~Values of the Lagrange multipliers $\omega_1$ and $\omega_2$
    corresponding to a marginal spectrum for multispherical spin glasses with
    $\sigma_1^2=f_1''(1)=1$, $\sigma_2^2=f_2''(1)=1$, and various $\epsilon$.
    \textbf{(b)}~Spectra corresponding to the parameters $\omega_1$ and
    $\omega_2$ marked by the circles in panel (a).
    \textbf{(c)}~The complexity of marginal minima in a multispherical model with
    $f_1(q)=\frac16q^3$ and $f_2(q)=\frac1{12}q^4$ for a variety of
    $\epsilon$. Since $f_1''(1)=f_2''(1)=1$, the marginal values correspond
    precisely to those in (a--b).
  } \label{fig:msg.marg}
\end{figure}

When the $\beta$-dependence of the $q$ variables is inserted into the effective
action \eqref{eq:multispherical.ansatz} and the limit $\beta\to\infty$ taken, we find an expression that is too large to report
here. However, it can be extremized over all of the variables in the problem
just as in the previous examples to find the values of the Lagrange multipliers
$\omega_1$ and $\omega_2$ corresponding to marginal minima.
Fig.~\ref{fig:msg.marg}(a) shows examples of the $\omega_1$ and
$\omega_2$ corresponding to marginal spectra for a variety of couplings $\epsilon$ when the covariances of the
energy on the two spherical subspaces are such that $1=f_1''(1)=f_2''(1)$.
Fig.~\ref{fig:msg.marg}(b) shows the Hessian spectra associated with some
specific pairs $(\omega_1,\omega_2)$. When $\epsilon=0$ and the two spheres are
uncoupled, we find the result for two independent spherical spin glasses: if
either $\omega_1=2\sqrt{f''(1)}=2$ or $\omega_2=2\sqrt{f''(1)}=2$ and the other
Lagrange multiplier is larger than 2, then we have a marginal minimum made up of the Cartesian product of a marginal minimum on one
subspace and a stable minimum on the other.

Fig.~\ref{fig:msg.marg}(c) shows the complexity of marginal minima in an
example where both $H_1$ and $H_2$ correspond to pure $p$-spin models, with
$f_1(q)=\frac16q^3$ and $f_2(q)=\frac1{12}q^4$. Despite having different
covariance functions, these both satisfy $1=f_1''(1)=f_2''(1)$ and therefore
have marginal minima for Lagrange multipliers that satisfy the relationships in
Fig.\ref{fig:msg.marg}(a). In the uncoupled system with $\epsilon=0$, the most
common type of marginal stationary point consists of independently marginal
stationary points in the two subsystems, with $\omega_1=\omega_2=2$. As
$\epsilon$ is increased, the most common type of marginal minimum drifts toward
points with $\omega_1>\omega_2$.

Multispherical spin glasses may be an interesting platform for testing ideas
about which among the possible marginal minima can attract dynamics
and which cannot. In the limit where $\epsilon=0$ and the configurations of the
two spheres are independent, the minima found dynamically should be marginal on both
subspaces. Just because technically on the expanded configuration space
the Cartesian product of a deep stable minimum on one sphere and a marginal minimum on the other is
a marginal minimum on the whole space doesn't mean the deep and stable minimum
is any easier to find. This intuitive idea that is precise in the zero-coupling
limit should continue to hold at small nonzero coupling, and perhaps reveal
something about the inherent properties of marginal minima that do not tend to be found
by algorithms.

\subsection{Sums of squared random functions}
\label{sec:least.squares}

In this subsection we consider perhaps the simplest example of a non-Gaussian
landscape: the problem of sums of squared random functions. This problem has a close resemblance to nonlinear least squares optimization. Though,
for reasons we will see it is easier to make predictions for nonlinear
\emph{most} squares, i.e., the problem of maximizing the sum of squared terms.
We again take a spherical configuration space with $\mathbf x\in S^{N-1}$ and $0=g(\mathbf x)=\frac12(\|\mathbf x\|^2-N)$ as in the spherical spin glasses. The energy is built from a set
of $M=\alpha N$ random functions $V_k:\mathbf S^{N-1}\to\mathbb R$ that are
centered Gaussians with covariance
\begin{equation}
  \overline{V_i(\mathbf x)V_j(\mathbf x')}=\delta_{ij}f\left(\frac{\mathbf x\cdot\mathbf x'}N\right)
\end{equation}
Each of the $V_k$ is an independent spherical spin glass.
The total energy is minus the sum of squares of the $V_k$, or
\begin{equation} \label{eq:ls.hamiltonian}
  H(\mathbf x)=-\frac12\sum_{k=1}^MV_k(\mathbf x)^2
\end{equation}
The landscape complexity and large deviations of the ground state for the
least-squares version of this problem were recently studied in a linear
context, with $f(q)=\sigma^2+aq$ \cite{Fyodorov_2019_A, Fyodorov_2020_Counting,
Fyodorov_2022_Optimization, Vivo_2024_Random}. Some results on the ground state of the general
nonlinear problem can also be found in \cite{Tublin_2022_A}, and a solution to
the equilibrium problem can be found in \cite{Urbani_2023_A}.
Those works indicate that the low-lying minima of the least squares problem tend
to be either replica symmetric or full replica symmetry breaking. To avoid
either a trivial analysis or a very complex one, we instead focus on maximizing
the sum of squares, or minimizing \eqref{eq:ls.hamiltonian}.

The minima of \eqref{eq:ls.hamiltonian} have a more amenable structure
for study than the maxima, as they are typically described by a {\oldstylenums1}\textsc{rsb}-like structure. There is a
heuristic intuition for this: in the limit of $M\to1$, this problem is just minus the
square of a spherical spin glass landscape. The distribution and properties of
stationary points low and high in the spherical spin glass are not changed,
except that their energies are stretched and maxima are transformed into
minima. Therefore, the bottom of the landscape doesn't qualitatively change. The
top, however, consists of the zero-energy level set in the spherical spin
glass. This level set is well-connected, and so the highest states should also
be well connected and flat.

Focusing on the bottom of the landscape and therefore dealing with a {\oldstylenums1}\textsc{rsb}-like
problem makes our analysis easier. Algorithms will tend to be stuck in the ways
they are in hard optimization problems, and we will be able to
predict where. Therefore, we will study the most squares problem rather than
the least squares one. We calculate the complexity of minima of
\eqref{eq:ls.hamiltonian} in Appendix~\ref{sec:dominant.complexity}, which
corresponds to maximizing the sum of squares, under a replica symmetric ansatz
(which covers {\oldstylenums1}\textsc{rsb}-like problems) for arbitrary covariance $f$, and we
calculate the complexity of marginal minima in this section.

As in the previous sections, we used the method of Lagrange multipliers to analyse stationary points on the constrained configuration space. The Lagrangian and its associated gradient and Hessian are
\begin{equation}
  L(\mathbf x,\omega)
  =-\frac12\bigg(\sum_k^MV_k(\mathbf x)^2-\omega\big(\|\mathbf x\|^2-N\big)\bigg) \\
\end{equation}
\begin{align}
  &\nabla H(\mathbf x,\omega)
  =-\sum_k^MV_k(\mathbf x)\partial V_k(\mathbf x)+\omega\mathbf x
  \\
  &\begin{aligned}
    &\operatorname{Hess}H(\mathbf x,\omega) \\
    &\qquad=-\sum_k^M\left[\partial V_k(\mathbf x)\partial V_k(\mathbf x)
    -V_k(\mathbf x)\partial\partial V_k(\mathbf x)\right]+\omega I
  \end{aligned}
\end{align}
Unlike in the spherical and multispherical spin glasses, the value $\mu^*$
defined in \eqref{eq:mu.star} giving the typical value of
$\frac1N\operatorname{Tr}\partial\partial H$ is not always zero. Instead
$\mu^*=-f'(0)$, nonzero where there is a linear term in $V$. Fixing the trace
of the Hessian is therefore equivalent to constraining the value of the Lagrange multiplier $\omega=\mu+f'(0)$.

The derivation of the marginal complexity for this model is complicated, but
can be made schematically like that of the derivation of the equilibrium free
energy by use of superspace coordinates. Following the framework outlined in
Section~\ref{sec:superspace_kac-rice}, the replicated number of stationary
points conditioned on energy $E$, trace $\mu$, and minimum eigenvalue
$\lambda^*$ is given by
\begin{widetext}
\begin{equation}
  \begin{aligned}
    \mathcal N(E,\mu,\lambda^*)^n
    &=\int d\hat\beta\,d\hat\lambda\prod_{a=1}^n\lim_{m_a\to0}\prod_{\alpha=1}^{m_a}d\pmb\phi_a^\alpha
    \\
    &\qquad\times\exp\left\{
      \delta^{\alpha1}N(\hat\beta E+\hat\lambda\lambda^*)
      -\frac12\int d1\,d2\,\left[B^\alpha(1,2)\sum_{k=1}^MV_k(\pmb\phi_a^\alpha(1,2))^2
      -\big(\mu+f'(0)\big)\|\pmb\phi_a^\alpha(1,2)\|^2\right]
    \right\}
  \end{aligned}
\end{equation}
The first step to evaluate this expression is to linearize the dependence on
the random functions $V$. This is accomplished by inserting into the integral a
Dirac $\delta$ function fixing the value of the energy for each replica, or
\begin{equation}
  \delta\big(
    V_k(\pmb\phi_a^\alpha(1,2))-v_{ka}^\alpha(1,2)
  \big)
  =
  \int d\hat v_{ka}^\alpha\exp\left[
    i\int d1\,d2\,\hat v_{ka}^\alpha(1,2)
    \big(V_k(\pmb\phi_a^\alpha(1,2))-v_{ka}^\alpha(1,2)\big)
  \right]
\end{equation}
where we have introduced auxiliary superfields $\hat v$. With this inserted into the
integral, all other instances of $V$ are replaced by $v$, and the only
remaining dependence on the disorder is from the term $\hat vV$ arising from
the Fourier representation of the Dirac $\delta$ function. This term is linear
in $V$, and therefore the random functions can be averaged over to produce
\begin{equation}
  \overline{
    \exp\left[
      i\sum_k^M\sum_a^n\sum_\alpha^{m_a}\int d1\,d2\,\hat v_{ka}^\alpha(1,2)
      V_k(\pmb\phi_a^\alpha(1,2))
    \right]
  }
  =
  -\frac12\sum_{ab}^n\sum_{\alpha\gamma}^{m_a}\sum_k^M\int d1\,d2\,d3\,d4\,
  \hat v_{ka}^\alpha(1,2)f\big(\pmb\phi_a^\alpha(1,2)\cdot\pmb\phi_b^\gamma(3,4)\big)\hat v_{kb}^\gamma(3,4)
\end{equation}
\end{widetext}
The entire integrand is now factorized in the indices $k$ and quadratic in the
superfields $v$ and $\hat v$ with the kernel
\begin{equation}
  \begin{bmatrix}
    B^\alpha(1,2)\delta(1,3)\delta(2,4)\delta_{ab}\delta^{\alpha\gamma}
    & i\delta(1,3)\,\delta(2,4) \delta_{ab}\delta^{\alpha\gamma}\\
    i\delta(1,3)\,\delta(2,4) \delta_{ab}\delta^{\alpha\gamma}
    & f\big(\pmb\phi_a^\alpha(1,2)\cdot\pmb\phi_b^\gamma(3,4)\big)
  \end{bmatrix}
\end{equation}
The integration over $v$ and $\hat v$ results in a term in the effective action of the form
\begin{equation} \label{eq:sdet.1}
  \begin{aligned}
    &-\frac M2\log\operatorname{sdet}\bigg[
      \delta(1,3)\,\delta(2,4) \delta_{ab}\delta^{\alpha\gamma} \\
    &\hspace{7em}+B^\alpha(1,2)f\big(\pmb\phi_a^\alpha(1,2)\cdot\pmb\phi_b^\gamma(3,4)\big)
    \bigg]
  \end{aligned}
\end{equation}
When expanded, the supermatrix
$\pmb\phi_a^\alpha(1,2)\cdot\pmb\phi_b^\gamma(3,4)$ is constructed of the
scalar products of the real and Grassmann vectors that make up $\pmb\phi$. The
change of variables to these order parameters again results in the Jacobian of
\eqref{eq:coordinate.jacobian}, contributing
\begin{equation}
  \frac N2\log\det J-\frac N2\log\det G^2
\end{equation}
to the effective action.

Up to this point, the expressions are general and independent of a given
ansatz. However, we expect that the order parameters $X$ and $\hat X$ are zero,
since again we are in a setting with no signal or external field. Applying this ansatz here avoids a dramatically
more complicated expression for the effective action. We also will apply the ansatz that $Q_{ab}^{\alpha\gamma}$ is zero for $a\neq b$, which is equivalent to assuming that the soft
directions of typical pairs of stationary points are uncorrelated, and further
that $Q^{\alpha\gamma}=Q_{aa}^{\alpha\gamma}$ independently of the index $a$,
implying that correlations in the tangent space of typical stationary points
are the same.

Given this ansatz, taking the superdeterminant in
\eqref{eq:sdet.1} yields
\begin{widetext}
\begin{align}
    &-\frac M2\log\det\left\{
      \left[
        f'(C)\odot D-\hat\beta I+\left(R^{\circ2}-G^{\circ2}+I\sum_{\alpha\gamma}2(\delta^{\alpha1}\hat\lambda+\beta)(\delta^{\gamma1}\hat\lambda+\beta)(Q^{\alpha\gamma})^2\right)\odot f''(C)
      \right]f(C)
      +(I-R\odot f'(C))^2
    \right\} \notag \\
    &\hspace{16em}-n\frac M2\log\det\big[\delta_{\alpha\gamma}-2(\delta_{\alpha1}\hat\lambda+\beta)Q^{\alpha\gamma}\big]
    +M\log\det\big[I+G\odot f'(C)\big]
\end{align}
\end{widetext}
where once again $\odot$ is the Hadamard product and $A^{\circ n}$ gives the
Hadamard power of $A$. We can already see one substantive difference between
the structure of this problem and that of the spherical models: the effective
action in this case mixes the order parameters $G$ due to the Grassmann variables with the
ones $C$, $R$, and $D$ due to the other variables. Notice further that the dependence on $Q$ due to the marginal constraint is likewise no longer separable into its own term. This is the realization of
the fact that the Hessian is no longer independent of the energy
and gradient.

Now we have reduced the problem to an extremal one over the order parameters
$\hat\beta$, $\hat\lambda$, $C$, $R$, $D$, $G$, and $Q$, it is time to make an
ansatz for the form of order we expect to find. We will focus on a regime where
the structure of stationary points is replica symmetric, and further where
typical pairs of stationary points have no overlap. This requires that $f(0)=0$, or that there is no constant term in the random functions. This gives the ansatz
\begin{align}
  C=I && R=rI && D = dI && G = gI
\end{align}
We further take a planted replica symmetric structure for the matrix $Q$,
identical to that in \eqref{eq:Q.structure}. This results in
\begin{equation}
  \begin{aligned}
    &\Sigma_{\lambda^*}(E,\mu)
    =\frac1N\lim_{n\to0}\frac\partial{\partial n}
    \int d\hat\beta\,d\hat\lambda\,dr\,dd\,dg\,dq_0\,d\tilde q_0\,\\
    &\hspace{8em}\times e^{nN\mathcal S_\mathrm{RSS}(\hat\beta,\hat\lambda,r,d,g,q_0,\tilde q_0\mid\lambda^*,E,\mu,\beta)}
  \end{aligned}
\end{equation}
with an effective action
\begin{widetext}
\begin{equation}
  \begin{aligned}
    &\mathcal S_\mathrm{RSS}(\hat\beta,\hat\lambda,r,d,g,q_0,\tilde q_0\mid\lambda^*,E,\mu,\beta)
    =\hat\beta E-\big(\mu+f'(0)\big)(r+g+\hat\lambda)
    +\hat\lambda\lambda^*
    +\frac12\log\left(\frac{d+r^2}{g^2}
    \times\frac{1-2q_0+\tilde q_0^2}{(1-q_0)^2}\right) \\
    &\quad-\frac\alpha2\log\Bigg(
      \frac{1-4f'(1)\big[\beta(1-q_0)+\frac12\hat\lambda-\beta(\beta+\hat\lambda)(1-2q_0+\tilde q_0^2)f'(1)\big]}
      {\big[1-2(1-q_0)\beta f'(1)\big]^2} \\
    &\qquad\qquad\qquad\times
      \frac{
        f(1)\big[f'(1)d-\hat\beta-f''(1)\big(r^2-g^2+4q_0^2\beta^2-4\tilde q_0^2\beta(\beta+\hat\lambda)+4\beta\hat\lambda+2\hat\lambda^2\big)\big]+(1-rf'(1))^2
      }{
        \big[1+gf'(1)\big]^2
      }
    \Bigg)
  \end{aligned}
\end{equation}
We expect as before the limits of $q_0$ and $\tilde q_0$ as $\beta$ goes to
infinity to approach one, defining their asymptotic expansion like in
\eqref{eq:q0.limit} and \eqref{eq:q0t.limit}. Upon making this substitution and
taking the zero-temperature limit, we find
\begin{equation}
  \begin{aligned}
    &\mathcal S_\mathrm{RSS}(\hat\beta,\hat\lambda,r,d,g,y,\Delta z\mid\lambda^*,E,\mu,\infty)
    =\hat\beta E-\big(\mu+f'(0)\big)(r+g+\hat\lambda)
    +\hat\lambda\lambda^*
    +\frac12\log\left(\frac{d+r^2}{g^2}\times\frac{y^2-2\Delta z}{y^2}\right)
    \\
    &-\frac\alpha2\log\left(
      \frac{
        1-2(2y+\hat\lambda)f'(1)+4(y^2-2\Delta z)f'(1)^2
      }{\big[1-2yf'(1)\big]^2}
      \times
      \frac{
        f(1)\big[f'(1)d-\hat\beta-f''(1)(r^2-g^2+8(y\hat\lambda+\Delta z)+2\hat\lambda^2)\big]+\big[1-rf'(1)\big]^2
      }{
        \big[1+gf'(1)\big]^2
      }
    \right)
  \end{aligned}
\end{equation}
\end{widetext}
We can finally write the complexity with fixed energy $E$, stability $\mu$, and
minimum eigenvalue $\lambda^*$ as
\begin{equation} \label{eq:rss.complexity}
  \begin{aligned}
    &\Sigma_{\lambda^*}(E,\mu) \\
    &=\underset{\hat\beta,\hat\lambda,r,d,g,y,\Delta z}{\operatorname{extremum}}
    \mathcal S_\textrm{RSS}(\hat\beta,\hat\lambda,r,d,g,y,\Delta z\mid\lambda^*,E,\mu,\infty)
  \end{aligned}
\end{equation}
Note that unlike the previous two examples, the effective action in this case
does not split into two largely independent pieces, one relating to the
eigenvalue problem and one relating to the ordinary complexity. Instead, the
order parameters related to the eigenvalue problem are mixed throughout the
effective action with those of the ordinary complexity. This is a signal of the
fact that the sum of squares problem is not Gaussian, while the previous two
examples are. In all non-Gaussian problems, conditioning on properties of the
Hessian cannot be done independently from the complexity, and the method
introduced in this paper becomes necessary.

\begin{figure}
  \includegraphics{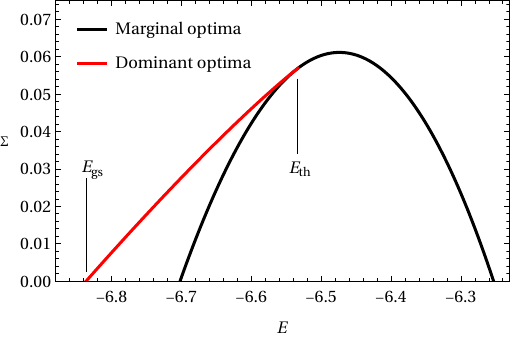}
  \caption{
    Dominant and marginal complexity in the nonlinear sum of squares problem
    for $\alpha=\frac32$ and $f(q)=q^2+q^3$. The ground state energy
    $E_\mathrm{gs}$ and the threshold energy $E_\mathrm{th}$ are marked on the
    plot.
  } \label{fig:ls.complexity}
\end{figure}

The marginal complexity can be derived from \eqref{eq:rss.complexity} using the condition \eqref{eq:marginal.stability} to fix $\mu$ to the marginal stability $\mu_\textrm m(E)$ and then evaluating the complexity at that stability as in \eqref{eq:marginal.complexity}.
Fig.~\ref{fig:ls.complexity} shows the marginal complexity in a sum-of-squares
model with $\alpha=\frac32$ and $f(q)=q^2+q^3$. Also shown is the dominant
complexity computed in Appendix~\ref{sec:dominant.complexity}. As the figure
demonstrates, the range of energies at which marginal minima are found can
differ significantly from those implied by the dominant complexity, with the
lowest energy significantly higher than the ground state and the highest energy
significantly higher than the threshold.

\begin{figure}
  \includegraphics{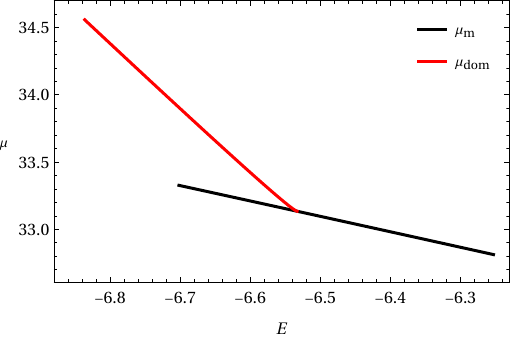}
  \caption{
    The stability, or shift of the trace, for dominant and marginal optima in
    the nonlinear sum of squares problem for $\alpha=\frac32$ and
    $f(q)=q^2+q^3$.
  } \label{fig:ls.stability}
\end{figure}

Fig.~\ref{fig:ls.stability} shows the associated marginal stability
$\mu_\mathrm m(E)$ for the same model. Recall that the definition of the
marginal stability in \eqref{eq:marginal.stability} is that which eliminates
the variation of $\Sigma_{\lambda^*}(E,\mu)$ with respect to $\lambda^*$ at the
point $\lambda^*=0$. Unlike in the Gaussian spherical spin glass, in this model $\mu_\mathrm m(E)$ varies with
energy in a nontrivial way. The figure also shows the dominant stability,
which is the stability associated with the dominant complexity and coincides
with the marginal stability only at the threshold energy.

\begin{figure}
  \includegraphics{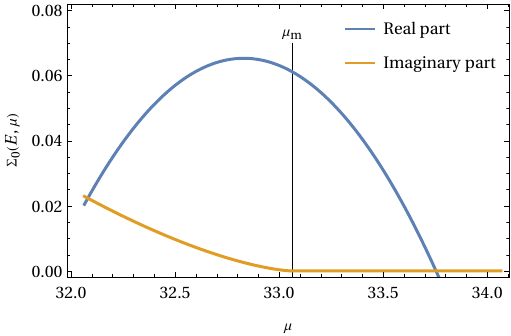}
  \caption{
    Real and imaginary parts of the complexity $\Sigma_0(E,\mu)$ with fixed
    minimum eigenvalue $\lambda^*=0$ as a function of $\mu$ in the nonlinear
    sum of squares problem with $\alpha=\frac32$, $f(q)=q^2+q^3$, and
    $E\simeq-6.47$. The vertical line depicts the value of the marginal
    stability $\mu_\mathrm m$.
  } \label{fig:ls.reim}
\end{figure}

Because this version of the model has no signal, we were able to use the heuristic
\eqref{eq:marginal.stability} to fix the marginal stability. However, we could
also have used the more general method for finding a pseudogapped Hessian
spectrum by locating the value of $\mu$ at which the complexity develops an
imaginary part, as described in Section \ref{sec:pseudogap} and pictured in
Fig.~\ref{fig:large.dev}. The real and imaginary parts of the complexity
$\Sigma_0(E,\mu)$ are plotted in Fig.~\ref{fig:ls.reim} as a function of $\mu$
at fixed energy. The figure also shows the marginal stability $\mu_\mathrm m$
predicted by the variational approach \eqref{eq:marginal.stability}. The
marginal stability corresponds to precisely the point at which an imaginary
part develops in the complexity. This demonstrates that the principles we used
to determine the marginal stability continue to hold even in non-Gaussian cases
where the complexity and the condition to fix the minimum eigenvalue are
tangled together.

In a related paper, we use a sum of squared random functions model to explore the relationship between the marginal
complexity and the performance of two generic algorithms:
gradient descent and approximate message passing
\cite{Kent-Dobias_2024_Algorithm-independent}. We show that the range of
energies where the marginal complexity is positive does effectively bound the
performance of these algorithms. At the moment the comparison is restricted to
models with small polynomial powers appearing in $f(q)$ and with small $\alpha$
for computational reasons. However, using the \textsc{dmft} results
already found for these models it should be possible to make comparisons in a
wider family of models \cite{Kamali_2023_Dynamical, Kamali_2023_Stochastic}.

The results for the marginal complexity are complimentary to rigorous results on
the performance of algorithms in the least squares case, which focus on bounds
for $\alpha$ and the parameters of $f$ necessary for zero-energy solutions to exist and be
found by algorithms \cite{Montanari_2023_Solving, Montanari_2024_On}. After more
work to evaluate the marginal complexity in the full \textsc{rsb} case, it will
be interesting to compare the bounds implied by the distribution of marginal
minima with those made by other means.

\section{Conclusions}
\label{sec:conclusion}

We have introduced a method for conditioning complexity on the marginality of
stationary points. This method is general, and permits
conditioning without first needing to understand the statistics of the Hessian at stationary points. We used our approach to study marginal complexity in three
different models of random landscapes, showing that the method works and can be
applied to models whose marginal complexity was not previously known. In related work, we further show that marginal complexity in the third
model of sums of squared random functions can be used to effectively bound
algorithmic performance \cite{Kent-Dobias_2024_Algorithm-independent}.

There are some limitations to the approach we relied on in this
paper. The main limitation is our restriction to signalless landscapes, where
there is no symmetry-breaking favored direction. This allowed us to treat stationary points with isolated eigenvalues as atypical, and
therefore find the marginal stability $\mu_\mathrm m$ using a variational principle.
However, most models of interest in inference have a nonzero signal strength
and therefore often have typical stationary points with an isolated eigenvalue.
As we described, marginal complexity can still be analyzed in these
systems by tuning the shift $\mu$ until the large-deviation principle breaks
down and an imaginary part of the complexity appears. However, this is an
inconvenient approach. It is possible that a variational approach can be
preserved by treating the direction toward and the directions orthogonal to the
signal differently. This problem merits further research.

Finally, the problem of predicting which marginal minima are able to attract
some dynamics and which cannot attract any dynamics looms large over this work.
As we discussed briefly at the end of Section~\ref{sec:multispherical}, in some
simple contexts it is easy to see why certain marginal minima are not viable,
but at the moment we do not know how to generalize this. Ideas related to the
self-similarity and stochastic stability of minima have
recently been suggested as a route to understanding this problem, but this approach is
still in its infancy \cite{Urbani_2024_Statistical}.

The title of our paper and that of \citeauthor{Muller_2006_Marginal} suggest
they address the same topic, but this is not the case
\cite{Muller_2006_Marginal}. That work differs in three important and
fundamental ways. First, it describes minima of the TAP free energy and
involves peculiarities specific to the TAP. Second, it describes dominant
minima which happen to be marginal, not a condition for finding subdominant marginal minima. Finally, it
focuses on minima with a single soft direction (which are the typical minima of
the low temperature Sherrington--Kirkpatrick TAP free energy), while we aim to
avoid such minima in favor of ones that have a pseudogap (which we argue are relevant
to out-of-equilibrium dynamics). The fact that the typical minima studied by
\citeauthor{Muller_2006_Marginal} are not marginal in this latter sense may
provide an intuitive explanation for the seeming discrepancy between the proof
that the low-energy Sherrington--Kirkpatrick model cannot be sampled
\cite{ElAlaoui_2022_Sampling} and the proof that a message passing algorithm
can find near-ground states \cite{Montanari_2021_Optimization}: the algorithm
finds the atypical low-lying states that are marginal in the sense considered
here but cannot find the typical ones considered by
\citeauthor{Muller_2006_Marginal}.

\begin{acknowledgements}
  JK-D is supported by a \textsc{DynSysMath} Specific Initiative of the INFN.
\end{acknowledgements}

\appendix

\section{A primer on superspace}
\label{sec:superspace}

In this appendix we review the algebra of superspace \cite{DeWitt_1992_Supermanifolds}.
The superspace $\mathbb R^{N|2D}$ is a vector space with $N$ real indices and
$2D$ Grassmann indices $\bar\theta_1,\theta_1,\ldots,\bar\theta_D,\theta_D$.
The Grassmann indices anticommute like fermions. Their integration is defined by
\begin{equation}
  \int d\theta\,\theta=1
  \qquad
  \int d\theta\,1=0
\end{equation}
Because the Grassmann indices anticommute, their square is always zero.
Therefore, any series expansion of a function with respect to a given Grassmann
index will terminate exactly at linear order, while a series expansion with
respect to $n$ Grassmann variables will terminate exactly at $n$th order. If
$f$ is an arbitrary superspace function, then the integral of $f$ with respect
to a Grassmann index can be evaluated using this property of the series
expansion by
\begin{equation}
  \int d\theta\,f(a+b\theta)
  =\int d\theta\,\left[f(a)+f'(a)b\theta\right]
  =f'(a)b
\end{equation}
This kind of behavior of integrals over the Grassmann indices makes them useful
for compactly expressing the Kac--Rice measure. To see why, consider the
specific superspace $\mathbb R^{N|2}$, where an arbitrary vector can be expressed as
\begin{equation}
  \pmb\phi(1)=\mathbf x+\bar\theta_1\pmb\eta+\bar{\pmb\eta}\theta_1+\bar\theta_1\theta_1i\hat{\mathbf x}
\end{equation}
where $\mathbf x,\hat{\mathbf x}\in\mathbb R^N$ and $\bar{\pmb\eta},\pmb\eta$ are
$N$-dimensional Grassmann vectors. The dependence of $\pmb\phi$ on 1 indicates
the index of Grassmann variables $\bar\theta_1,\theta_1$ inside, since we will
sometimes want to use, e.g., $\pmb\phi(2)$ defined identically save for
substitution by $\bar\theta_2,\theta_2$. Consider the series expansion of an arbitrary function $f$ of this supervector:
\begin{equation}
  \begin{aligned}
    f\big(\pmb\phi(1)\big)
    &=f(\mathbf x)
    +\big(\bar\theta_1\pmb\eta+\bar{\pmb\eta}\theta_1+\bar\theta_1\theta_1i\hat{\mathbf x}\big)^T\partial f(\mathbf x) \\
    &\quad+\frac12\big(\bar\theta_1\pmb\eta+\bar{\pmb\eta}\theta_1\big)^T\partial\partial f(\mathbf x)\big(\bar\theta_1\pmb\eta+\bar{\pmb\eta}\theta_1\big) \\
    &=f(\mathbf x)
    +\big(\bar\theta_1\pmb\eta+\bar{\pmb\eta}\theta_1+\bar\theta_1\theta_1i\hat{\mathbf x}\big)^T\partial f(\mathbf x) \\
    &\qquad-\bar\theta_1\theta_1\bar{\pmb\eta}^T\partial\partial f(\mathbf x)\pmb\eta
  \end{aligned}
\end{equation}
where the last step we used the fact that the Hessian matrix is symmetric and
that squares of Grassmann indicies vanish. Using the integration rules defined above, we find
\begin{equation}
  \int d\theta_1\,d\bar\theta_1\,f\big(\pmb\phi(1)\big)
  =i\hat{\mathbf x}^T\partial f(\mathbf x)-\bar{\pmb\eta}^T\partial\partial f(\mathbf x)\pmb\eta
\end{equation}
These two terms are precisely the exponential representation of the Dirac
$\delta$ function of the gradient and determinant of the Hessian (without
absolute value sign) that make up the basic Kac--Rice measure, so that we can write
\begin{equation}
  \begin{aligned}
    &\int d\mathbf x\,\delta\big(\nabla H(\mathbf x)\big)\,\det\operatorname{Hess}H(\mathbf x) \\
    &\qquad=\int d\mathbf x\,d\bar{\pmb\eta}\,d\pmb\eta\,\frac{d\hat{\mathbf x}}{(2\pi)^N}\,e^{i\hat{\mathbf x}^T\nabla H(\mathbf x)-\bar{\pmb\eta}^T\operatorname{Hess}H(\mathbf x)\pmb\eta} \\
    &\qquad=\int d\pmb\phi\,e^{\int d1\,H(\pmb\phi(1))}
  \end{aligned}
\end{equation}
where we have written the measures $d1=d\theta_1\,d\bar\theta_1$ and $d\pmb\phi=d\mathbf
x\,d\bar{\pmb\eta}\,d\pmb\eta\,\frac{d\hat{\mathbf x}}{(2\pi)^N}$. Besides some
deep connections to the physics of BRST, this compact notation dramatically
simplifies the analytical treatment of the problem. The energy of stationary points can also be fixed using this notation, by writing
\begin{equation}
  \int d\pmb\phi\,d\hat\beta\,e^{\hat\beta E+\int d1\,(1-\hat\beta\bar\theta_1\theta_1)H(\pmb\phi(1))}
\end{equation}
which a small calculation confirms results in the same expression as \eqref{eq:delta.energy}.

The reason why this transformation is a simplification is because there are a large variety of superspace algebraic and
integral operations with direct corollaries to their ordinary real
counterparts. For instance, consider a super linear operator $M(1,2)$, which
like the super vector $\pmb\phi$ is made up of a linear combination of $N\times
N$ regular or Grassmann matrices indexed by every nonvanishing combination of
the Grassmann indices $\bar\theta_1,\theta_1,\bar\theta_2,\theta_2$. Such a supermatrix acts on supervectors by ordinary matrix multiplication and convolution in the Grassmann indices, i.e.,
\begin{equation}
  (M\pmb\phi)(1)=\int d2\,M(1,2)\pmb\phi(2)
\end{equation}
The identity supermatrix is given by
\begin{equation}
  \delta(1,2)=(\bar\theta_1-\bar\theta_2)(\theta_1-\theta_2)I
\end{equation}
Integrals involving superfields contracted into such operators result in schematically familiar expressions, like that of the standard Gaussian:
\begin{equation}
  \int d\pmb\phi\,e^{-\frac12\int\,d1\,d2\,\pmb\phi(1)^TM(1,2)\pmb\phi(2)}
  =(\operatorname{sdet}M)^{-1/2}
\end{equation}
where the usual role of the determinant is replaced by the superdeterminant.
The superdeterminant can be defined using the ordinary determinant by writing a
block version of the matrix $M$. If $\mathbf e(1)=\{1,\bar\theta_1\theta_1\}$ is
the basis vector of the even subspace of the superspace and $\mathbf
f(1)=\{\bar\theta_1,\theta_1\}$ is that of the odd subspace, dual bases $\mathbf e^\dagger(1)=\{\bar\theta_1\theta_1,1\}$ and $\mathbf f^\dagger(1)=\{-\theta_1,\bar\theta_1\}$ can be defined by the requirement that
\begin{align}
  &\int d1\,e_i^\dagger(1)e_j(1)=\delta_{ij}
  &&
  \int d1\,f_i^\dagger(1)f_j(1)=\delta_{ij} \\
  &\int d1\,e_i^\dagger(1)f_j(1)=0
  &&
  \int d1\,f_i^\dagger(1)e_j(1)=0
\end{align}
With such bases and dual bases defined, we can form a
block representation of $M$ in analogy to the matrix form of an operator in quantum mechanics by
\begin{equation}
  \int d1\,d2\,\begin{bmatrix}
    \mathbf e^\dagger(1)M(1,2)\mathbf e(2)
    &
    \mathbf e^\dagger(1)M(1,2)\mathbf f(2)
    \\
    \mathbf f^\dagger(1)M(1,2)\mathbf e(2)
    &
    \mathbf f^\dagger(1)M(1,2)\mathbf f(2)
  \end{bmatrix}
  =\begin{bmatrix}
    A & B \\ C & D
  \end{bmatrix}
\end{equation}
where each of the blocks is a $2N\times 2N$ real matrix. Then the
superdeterminant of $M$ is given by
\begin{equation}
  \operatorname{sdet}M=\det(A-BD^{-1}C)\det(D)^{-1}
\end{equation}
which is the same as the normal expression for the determinant of a block matrix
save for the inverse of $\det D$. Likewise, the supertrace of $M$ is is given by
\begin{equation}
  \operatorname{sTr}M=\operatorname{Tr}A-\operatorname{Tr}D
\end{equation}
The same method can be used to calculate the
superdeterminant and supertrace in arbitrary superspaces, where for $\mathbb R^{N|2D}$ each
basis has $2^{2D-1}$ elements. For instance, for $\mathbb R^{N|4}$ we have
\begin{align}
  &\mathbf e(1,2)=\{
    1,\bar\theta_1\theta_1,\bar\theta_2\theta_2,
    \bar\theta_1\theta_2,\bar\theta_2\theta_1,
    \bar\theta_1\bar\theta_2,\theta_1\theta_2,
    \bar\theta_1\theta_1\bar\theta_2\theta_2
  \}\notag \\
  &\mathbf f(1,2)=\{
    \bar\theta_1,\theta_1,\bar\theta_2,\theta_2,
    \bar\theta_1\theta_1\bar\theta_2,\bar\theta_2\theta_2\theta_1,
    \bar\theta_1\theta_1\theta_2,\bar\theta_2\theta_2\theta_1
  \}
\end{align}
with the dual bases defined analogously to those above.

\section{BRST symmetry}
\label{sec:brst}

When the trace $\mu$ is not fixed, there is an unusual symmetry in the dominant
complexity of minima \cite{Annibale_2003_The, Annibale_2003_Supersymmetric, Annibale_2004_Coexistence}.
This arises from considering the Kac--Rice formula as a kind of gauge fixing
procedure \cite{Zinn-Justin_2002_Quantum}. Around each stationary point
consider making the coordinate transformation $\mathbf u=\nabla H(\mathbf x)$.
Then, in the absence of fixing the trace of the Hessian to $\mu$, the Kac--Rice measure becomes
\begin{equation}
    \int d\nu(\mathbf x,\pmb\omega\mid E)
    =\int\sum_\sigma d\mathbf u\,\delta(\mathbf u)\,
    \delta\big(NE-H(\mathbf x_\sigma)\big)
\end{equation}
where the sum is over stationary points $\sigma$. This integral has a symmetry of its
measure of the form $\mathbf u\mapsto\mathbf u+\delta\mathbf u$. Under the
nonlinear transformation that connects $\mathbf u$ and $\mathbf x$, this
implies a symmetry of the measure in the Kac--Rice integral of $\mathbf
x\mapsto\mathbf x+(\operatorname{Hess}H)^{-1}\delta\mathbf u$. This symmetry, while exact, is
nonlinear and difficult to work with.

When the absolute value function has been dropped and Grassmann vectors introduced to represent the determinant of the Hessian,
this symmetry can be simplified considerably. Due to the expansion properties
of Grassmann integrals, any appearance of $-\bar{\pmb\eta}\pmb\eta^T$ in the
integrand resolves to $(\operatorname{Hess}H)^{-1}$. The
symmetry of the measure can then be written
\begin{equation}
  \mathbf x\mapsto \mathbf x-\bar{\pmb\eta}\pmb\eta^T\delta\mathbf u
  =\mathbf x+\bar{\pmb\eta}\delta\epsilon
\end{equation}
where $\delta\epsilon=-\pmb\eta^T\delta\mathbf u$ is a Grassmann number. This
establishes that $\delta\mathbf x=\bar{\pmb\eta}\delta\epsilon$, now linear. The rest of
the transformation can be built by requiring that the action is invariant after
expansion in $\delta\epsilon$. This gives
\begin{align}
  \delta\mathbf x=\bar{\pmb\eta}\,\delta\epsilon &&
  \delta\hat{\mathbf x}=-i\hat\beta\bar{\pmb\eta}\,\delta\epsilon &&
  \delta\pmb\eta=-i\hat{\mathbf x}\,\delta\epsilon  &&
  \delta\bar{\pmb\eta}=0
\end{align}
so that the differential form of the symmetry is
\begin{equation}
  \mathcal D=\bar{\pmb\eta}\cdot\frac\partial{\partial\mathbf x}
  -i\hat\beta\bar{\pmb\eta}\cdot\frac\partial{\partial\hat{\mathbf x}}
  -i\hat{\mathbf x}\cdot\frac\partial{\partial\pmb\eta}
\end{equation}
The Ward identities associated with this symmetry give rise to relationships
among the order parameters. These identities come from applying the
differential symmetry to Grassmann-valued order parameters, and are
\begin{align}
  \begin{aligned}
    0&=\frac1N\mathcal D\langle\mathbf x_a\cdot\pmb\eta_b\rangle
    =\frac1N\left[
      \langle\bar{\pmb\eta}_a\cdot\pmb\eta_b\rangle-
      i\langle\mathbf x_a\cdot\hat{\mathbf x}_b\rangle
    \right] \\
     &=G_{ab}+R_{ab}
  \end{aligned} \\
  \begin{aligned}
    0&=\frac iN\mathcal D\langle\hat{\mathbf x}_a\cdot\pmb\eta_b\rangle
    =\frac1N\left[
      \hat\beta\langle\bar{\pmb\eta}_a\cdot\pmb\eta_b\rangle
      +\langle\hat{\mathbf x}_a\cdot\hat{\mathbf x}_b\rangle
    \right] \\
     &=\hat\beta G_{ab}+D_{ab}
  \end{aligned}
\end{align}
These identities establish $G_{ab}=-R_{ab}$ and $D_{ab}=\hat\beta R_{ab}$,
allowing elimination of the matrices $G$ and $D$ in favor of $R$. Fixing the
trace to $\mu$ explicitly breaks this symmetry, and the simplification is lost.

\section{Spectral density in the multispherical spin glass}
\label{sec:multispherical.spectrum}

In this appendix we derive an expression for the asymptotic spectral density of the Hessian in
the two-sphere multispherical spin glass that we describe in Section
\ref{sec:multispherical}. We use a typical approach of employing replicas to
compute the resolvent \cite{Livan_2018_Introduction}. The resolvent for the
Hessian of the multispherical model is given by an integral over $\mathbf
y=[\mathbf y^{(1)},\mathbf y^{(2)}]\in\mathbb R^{2N}$ as
\begin{widetext}
\begin{equation}
  \begin{aligned}
    G(\lambda)
    &=\lim_{n\to0}\int\|\mathbf y_1\|^2\,\prod_{a=1}^nd\mathbf y_a\,
    \exp\left\{
      -\frac12\mathbf y_a^T(\operatorname{Hess}H(\mathbf x,\pmb\omega)-\lambda I)\mathbf y_a
    \right\} \\
    &
    =\lim_{n\to0}\int\big(\|\mathbf y_1^{(1)}\|^2+\|\mathbf y_1^{(2)}\|^2\big)\,\prod_{a=1}^nd\mathbf y_a\,
    \exp\left\{
      -\frac12\begin{bmatrix}\mathbf y_a^{(1)}\\\mathbf y_a^{(2)}\end{bmatrix}^T
      \left(
        \begin{bmatrix}
          \partial_1\partial_1H_1(\mathbf x^{(1)})+\omega_1I & -\epsilon I \\
          -\epsilon I & \partial_2\partial_2H_2(\mathbf x^{(2)})+\omega_2I
        \end{bmatrix}
        -\lambda I
      \right)\begin{bmatrix}\mathbf y_a^{(1)}\\\mathbf y_a^{(2)}\end{bmatrix}
    \right\}
  \end{aligned}
\end{equation}
If $Y_{ab}^{(ij)}=\frac1N\mathbf y_a^{(i)}\cdot\mathbf y_b^{(j)}$ is the matrix
of overlaps of the vectors $\mathbf y$, then a short and standard calculation involving the average over $H$ and the change of variables from $\mathbf y$ to $Y$ yields
\begin{equation}
  \overline{G(\lambda)}=N\lim_{n\to0}\int dY\,\big(Y_{11}^{(11)}+Y_{11}^{(22)}\big)\,
  e^{nN\mathcal S(Y)}
\end{equation}
where the effective action $\mathcal S$ is given by
\begin{equation}
  \begin{aligned}
    \mathcal S(Y)
    =\lim_{n\to0}\frac1n\Bigg\{
      \frac14\sum_{ab}^n\left[
      f_1''(1)(Y_{ab}^{(11)})^2
      +f_2''(1)(Y_{ab}^{(22)})^2
    \right]
    +\frac12\sum_a^n\left[
      2\epsilon Y_{aa}^{(12)}
      +(\lambda-\omega_1)Y_{aa}^{(11)}
      +(\lambda-\omega_2)Y_{aa}^{(22)}
    \right] \qquad\\
    +\frac12\log\det\begin{bmatrix}
      Y^{(11)}&Y^{(12)}\\Y^{(12)}&Y^{(22)}
    \end{bmatrix}
    \Bigg\}
  \end{aligned}
\end{equation}
\end{widetext}
Making the replica symmetric ansatz $Y_{ab}^{(ij)}=y^{(ij)}\delta_{ab}$ for
each of the matrices $Y^{(ij)}$ yields
\begin{equation}
  \begin{aligned}
    \mathcal S(y)
    &=
      \frac14\left[f_1''(1)(y^{(11)})^2
      +f_2''(1)(y^{(22)})^2\right]+\epsilon y^{(12)}
      \\
    &
    \qquad+\frac12\left[(\lambda-\omega_1)y^{(11)}
    +(\lambda-\omega_2)y^{(22)}\right] \\
    &
    \qquad+\frac12\log(
      y^{(11)}y^{(22)}-y^{(12)}y^{(12)}
      )
  \end{aligned}
\end{equation}
while the average resolvent becomes
\begin{equation}
  \overline{G(\lambda)}
  =N(y^{(11)}+y^{(22)})
\end{equation}
for $y^{(11)}$ and $y^{(22)}$ evaluated at a saddle point of $\mathcal S$. The
spectral density at large $N$ is then given by the discontinuity in its
imaginary point on the real axis, or
\begin{equation}
  \rho(\lambda)
  =\frac1{2\pi iN}
  \left(
    \overline{G(\lambda+i0^+)}-\overline{G(\lambda+i0^-)}
  \right)
\end{equation}

\section{Complexity of dominant optima for sums of squared random functions}
\label{sec:dominant.complexity}

Here we share an outline of the derivation of formulas for the complexity of
dominant optima in sums of squared random functions of section
\ref{sec:least.squares}. While in this paper we only treat problems with a
replica symmetric structure, formulas for the effective action are generic to
any \textsc{rsb} structure and provide a starting point for analyzing the challenging
full \textsc{rsb} setting.

Using the $\mathbb R^{N|2}$ superfields
\begin{equation}
  \pmb\phi_a(1)=\mathbf x_a+\bar\theta_1\pmb\eta_a+\bar{\pmb\eta}_a\theta_1+\bar\theta_1\theta_1\hat{\mathbf x}_a,
\end{equation}
the replicated count of stationary points can be written
\begin{equation}
  \begin{aligned}
    &\mathcal N(E,\mu)^n
    =\int d\hat\beta\prod_{a=1}^n\,d\pmb\phi_a\,
    \exp\bigg[
      N\hat\beta E \\
    &-\frac12\int d1\,\left(
        B(1)\sum_{k=1}^MV_k(\pmb\phi_a(1))^2
        -\big(\mu+f'(0)\big)\|\pmb\phi_a(1)\|^2
      \right)
    \bigg]
  \end{aligned}
\end{equation}
for $B(1)=1-\hat\beta\bar\theta_1\theta_1$.
The derivation of the complexity follows from here nearly identically to that
in Appendix A.2 of \citeauthor{Fyodorov_2022_Optimization} with superoperations
replacing standard ones \cite{Fyodorov_2022_Optimization}. First we insert
Dirac $\delta$ functions to fix each of the $M$ energies $V_k(\pmb\phi_a(1))$ as
\begin{equation} \label{eq:Vv.delta}
  \begin{aligned}
    &\delta\big(V_k(\pmb\phi_a(1))-v_{ka}(1)\big)
    \\
    &=\int d\hat v_{ka}\,\exp\left[i\int d1\,\hat v_{ka}(1)\big(V_k(\pmb\phi_a(1))-v_{ka}(1)\big)\right]
  \end{aligned}
\end{equation}
The squared $V_k$ appearing in the energy can now be replaced by the variables
$v_k$, leaving the only remaining dependence on the disordered $V$ in the
contribution of \eqref{eq:Vv.delta}, which is linear. The average over the
disorder can then be computed, which yields
\begin{equation}
  \begin{aligned}
    &\overline{\exp\left[i\sum_{k=1}^M\sum_{a=1}^n\int d1\,\hat v_{ka}(1)V_k(\pmb\phi_a(1))\right]}
    \\
    &
    =\exp\left[
      -\frac12\sum_{k=1}^M\sum_{ab}^n\int d1\,d2\,\hat v_{ka}(1)f\left(\frac{\pmb\phi_a(1)\cdot\pmb\phi_b(2)}N\right)\hat v_{kb}(2)
    \right]
  \end{aligned}
\end{equation}
The result is factorized in the indices $k$ and Gaussian in the superfields $v$
and $\hat v$ with kernel
\begin{equation}
  \begin{bmatrix}
    B(1)\delta_{ab}\delta(1,2) & i\delta_{ab}\delta(1,2) \\
    i\delta_{ab}\delta(1,2) & f\left(\frac{\pmb\phi_a(1)\cdot\pmb\phi_b(2)}N\right)
  \end{bmatrix}
\end{equation}
Making the $M$ independent Gaussian integrals, we find
\begin{equation}
  \begin{aligned}
    &\mathcal N(E,\mu)^n
    =\int d\hat\beta\left(\prod_{a=1}^nd\pmb\phi_a\right) \\
    &\times\exp\bigg[
      nN\hat\beta E+\frac{\mu+f'(0)}2\sum_a^n\int d1\,\|\pmb\phi_a\|^2 \\
    &\quad-\frac M2\log\operatorname{sdet}\left(
        \delta_{ab}\delta(1,2)+B(1)f\left(\frac{\pmb\phi_a(1)\cdot\pmb\phi_b(2)}N\right)
      \right)
    \bigg]
  \end{aligned}
\end{equation}
We make a change of variables from the fields $\pmb\phi$ to matrices $\mathbb
Q_{ab}(1,2)=\frac1N\pmb\phi_a(1)\cdot\pmb\phi_b(2)$. This transformation results
in a change of measure of the form
\begin{equation}
  \prod_{a=1}^n d\pmb\phi_a=d\mathbb Q\,(\operatorname{sdet}\mathbb Q)^\frac N2
  =d\mathbb Q\,\exp\left[\frac N2\log\operatorname{sdet}\mathbb Q\right]
\end{equation}
\begin{widetext}
We therefore have
\begin{equation}
  \begin{aligned}
    &\mathcal N(E,\mu)^n
    =\int d\hat\beta\,d\mathbb Q\,
    \exp\bigg\{
      nN\hat\beta E+N\frac{\mu+f'(0)}2\operatorname{sTr}\mathbb Q
    +\frac N2\log\operatorname{sdet}\mathbb Q
    -\frac M2\log\operatorname{sdet}\left[
      \delta_{ab}\delta(1,2)+B(1)f(\mathbb Q_{ab}(1,2))
    \right]
  \bigg\}
  \end{aligned}
\end{equation}
We now need to blow up our supermatrices into our physical order parameters. We
have from the definition of $\pmb\phi$ and $\mathbb Q$ that
\begin{equation}
  \begin{aligned}
    &\mathbb Q_{ab}(1,2)
    =C_{ab}-G_{ab}(\bar\theta_1\theta_2+\bar\theta_2\theta_1)
    -R_{ab}(\bar\theta_1\theta_1+\bar\theta_2\theta_2)
    -D_{ab}\bar\theta_1\theta_2\bar\theta_2\theta_2
  \end{aligned}
\end{equation}
where $C$, $R$, $D$, and $G$ are the matrices defined in
\eqref{eq:order.parameters}. Other possible combinations involving scalar
products between fermionic and bosonic variables do not contribute at physical
saddle points \cite{Kurchan_1992_Supersymmetry}. Inserting this expansion into
the expression above and evaluating the superdeterminants and supertrace, we find
\begin{equation}
  \mathcal N(E,\mu)^n=\int d\hat\beta\,dC\,dR\,dD\,dG\,e^{nN\mathcal S_\mathrm{KR}(\hat\beta,C,R,D,G)}
\end{equation}
where the effective action is given by
\begin{equation}
  \begin{aligned}
    \mathcal S_\mathrm{KR}(\hat\beta,C,R,D,G)
    &=\hat\beta E+\lim_{n\to0}\frac1n\Bigg(-\big(\mu+f'(0)\big)\operatorname{Tr}(G+R)
    +\frac12\log\det\big[G^{-2}(CD+R^2)\big]
    +\alpha\log\det\big[I+G\odot f'(C)\big]
    \\
    &\qquad-\frac\alpha2\log\det\left[
      \Big(
        f'(C)\odot D-\hat\beta I+(G^{\circ2}-R^{\circ2})\odot f''(C)
      \Big)f(C)
      +(I-R\odot f'(C))^2
    \right]\Bigg)
  \end{aligned}
\end{equation}
where $\odot$ gives the Hadamard or componentwise product between the matrices
and $A^{\circ n}$ gives the Hadamard power of $A$, while other products and
powers are matrix products and powers.

In the case where $\mu$ is not specified, we can make use of the BRST symmetry
of Appendix~\ref{sec:brst} whose Ward identities give $D=\hat\beta R$ and
$G=-R$. Using these relations, the effective action becomes particularly
simple:
\begin{equation}
  \mathcal S_\mathrm{KR}(\hat\beta, C, R)
  =
  \hat\beta E
  +\frac12\lim_{n\to0}\frac1n\Big(
    \log\det(I+\hat\beta CR^{-1})
    -\alpha\log\det\left[
      I-\hat\beta f(C)\big(I-R\odot f'(C)\big)^{-1}
    \right]
  \Big)
\end{equation}
This effective action is general for arbitrary matrices $C$ and $R$, and
therefore arbitrary \textsc{rsb} order. When using a replica symmetric ansatz
of $C_{ab}=\delta_{ab}+c_0(1-\delta_{ab})$ and
$R_{ab}=r\delta_{ab}+r_0(1-\delta_{ab})$, the resulting function of
$\hat\beta$, $c_0$, $r$, and $r_0$ is
\begin{equation}
  \begin{aligned}
    &\mathcal S_\mathrm{KR}(\hat\beta,c_0,r,r_0)=
    \hat\beta E
    +\frac12\left[
      \log\left(1+\frac{\hat\beta(1-c_0)}{r-r_0}\right)
      +\frac{\hat\beta c_0+r_0}{\hat\beta(1-c_0)+r-r_0}
      -\frac{r_0}{r-r_0}
    \right]
    \\
    &\qquad-\frac\alpha 2\left[
      \log\left(1-\frac{\hat\beta\big(f(1)-f(c_0)\big)}{1-rf'(1)+r_0f'(c_0)}\right)
      -\frac{\hat\beta f(c_0)+r_0f'(c_0)}{
        1-\hat\beta\big(f(1)-f(c_0)\big)-rf'(1)+rf'(c_0)
      }+\frac{r_0f'(c_0)}{1-rf'(1)+r_0f'(c_0)}
    \right]
  \end{aligned}
\end{equation}
When $f(0)=0$ as in the cases directly studied in this work, this further
simplifies as $c_0=r_0=0$. The effective action is then
\begin{equation}
  \mathcal S_\mathrm{KR}(\hat\beta,r)=
  \hat\beta E
  +\frac12
    \log\left(1+\frac{\hat\beta}{r}\right)
  -\frac\alpha 2
    \log\left(1-\frac{\hat\beta f(1)}{1-rf'(1)}\right)
\end{equation}
Extremizing this expression with respect to the
order parameters $\hat\beta$ and $r$ produces the red line of dominant minima
shown in Fig.~\ref{fig:ls.complexity}.
\end{widetext}

\bibliography{marginal}

\end{document}